\documentclass[apj]{emulateapj}
\usepackage{rotating}
\usepackage{lscape}

\slugcomment{To appear in The Astrophysical Journal}

\shorttitle{Star formation in AGN host galaxies}
\shortauthors{Silverman et al.}

\begin{document}


\title{Ongoing and co-evolving star formation in zCOSMOS galaxies
hosting Active Galactic Nuclei}


\author{
J.~D.~Silverman\altaffilmark{1},
F.~Lamareille\altaffilmark{5},
C.~Maier\altaffilmark{1},
S.~Lilly\altaffilmark{1},
V.~Mainieri\altaffilmark{3},
M.~Brusa\altaffilmark{2},
N.~Cappelluti\altaffilmark{2},
G.~Hasinger\altaffilmark{2},
G.~ Zamorani\altaffilmark{4},
M.~ Scodeggio\altaffilmark{8},
M.~Bolzonella\altaffilmark{4},
T.~Contini\altaffilmark{5},
C. M.~Carollo\altaffilmark{1},
K.~Jahnke\altaffilmark{19}
J.-P.~ Kneib\altaffilmark{7},
O.~ Le F\`{e}vre\altaffilmark{7},
A.~Merloni\altaffilmark{2,16},
S.~Bardelli\altaffilmark{4},
A.~Bongiorno\altaffilmark{2},
H.~Brunner\altaffilmark{2},
K.~Caputi\altaffilmark{1},
F.~Civano\altaffilmark{15},
A.~Comastri\altaffilmark{4},
G.~ Coppa\altaffilmark{4},
O.~ Cucciati\altaffilmark{6},
S.~ de la Torre\altaffilmark{7},
L.~ de Ravel\altaffilmark{7},
M.~Elvis\altaffilmark{15},
A.~Finoguenov\altaffilmark{2},
F.~Fiore\altaffilmark{13},
P.~ Franzetti\altaffilmark{8},
B.~ Garilli\altaffilmark{8},
R.~Gilli\altaffilmark{4},
A.~ Iovino\altaffilmark{6},
P.~ Kampczyk\altaffilmark{1},
C. ~Knobel\altaffilmark{1},
K.~ Kova\v{c}\altaffilmark{1},
J.-F.~ Le Borgne\altaffilmark{5},
V.~ Le Brun\altaffilmark{7},
M.~ Mignoli\altaffilmark{4},
R.~ Pello\altaffilmark{5},
Y.~ Peng\altaffilmark{1},
E.~ Perez Montero\altaffilmark{5},
E.~ Ricciardelli\altaffilmark{12},
M.~ Tanaka\altaffilmark{3},
L.~ Tasca\altaffilmark{7},
L.~ Tresse\altaffilmark{7},
D.~ Vergani\altaffilmark{4},
C.~Vignali,\altaffilmark{9},
E.~ Zucca\altaffilmark{4},
D.~ Bottini\altaffilmark{8},
A.~ Cappi\altaffilmark{4},
P.~ Cassata\altaffilmark{7},
M.~ Fumana\altaffilmark{8},
R.~Griffiths\altaffilmark{18},
J.~ Kartaltepe\altaffilmark{20},
C.~ Marinoni\altaffilmark{10},
H.~ J. McCracken\altaffilmark{11},
P.~ Memeo\altaffilmark{8},
B.~ Meneux\altaffilmark{2,17},
P.~ Oesch\altaffilmark{1},
C.~ Porciani\altaffilmark{1},
M.~Salvato\altaffilmark{14},
}


\altaffiltext{1}{Institute of Astronomy, Swiss Federal Institute of Technology (ETH H\"onggerberg), CH-8093, Z\"urich, Switzerland.}
\altaffiltext{2}{Max-Planck-Institut f\"ur extraterrestrische Physik, D-84571 Garching, Germany}
\altaffiltext{3}{European Southern Observatory, Karl-Schwarzschild-Strasse 2, Garching, D-85748, Germany}
\altaffiltext{4}{INAF Osservatorio Astronomico di Bologna, via Ranzani 1, I-40127, Bologna, Italy}
\altaffiltext{5}{Laboratoire d'Astrophysique de Toulouse-Tarbes, Universit\'{e} de Toulouse, CNRS, 14 avenue Edouard Belin, F-31400 Toulouse, France}
\altaffiltext{6}{INAF Osservatorio Astronomico di Brera, Milan, Italy}
\altaffiltext{7}{Laboratoire d'Astrophysique de Marseille, Marseille, France}
\altaffiltext{8}{INAF - IASF Milano, Milan, Italy}
\altaffiltext{9}{Dipartimento di Astronomia, Universit\'a di Bologna, via Ranzani 1, I-40127, Bologna, Italy}
\altaffiltext{10}{Centre de Physique Theorique, Marseille, Marseille, France}
\altaffiltext{11}{Institut d'Astrophysique de Paris, UMR 7095 CNRS, Universit\'e Pierre et Marie Curie, 98 bis Boulevard Arago, F-75014 Paris, France.}
\altaffiltext{12}{Dipartimento di Astronomia, Universita di Padova, Padova, Italy}
\altaffiltext{13}{INAF, Osservatorio di Roma, Monteporzio Catone (RM), Italy}
\altaffiltext{14}{California Institute of Technology, Pasadena, CA, USA.}
\altaffiltext{15}{Harvard-Smithsonian Center for Astrophysics, 60 Garden Street, Cambridge, MA, 02138}
\altaffiltext{16}{Excellence Cluster Universe, Boltzmannstrasse 2, D-85748, Garching beu Muenchen, Germany}
\altaffiltext{17}{Universitats-Sternwarte, Scheinerstrasse 1, D-81679 Muenchen, Germany}
\altaffiltext{18}{Department of Physics, Carnegie Mellon University, 5000 Forbes Avenue, Pittsburgh, PA 15213}
\altaffiltext{19}{Max-Planck-Institut f\"ur Astronomie, K\"onigstuhl 17, D-69117 Heidelberg, Germany}
\altaffiltext{20}{Institute for Astronomy, University of Hawaii, 2680 Woodlawn Drive, Honolulu, HI, 96822}


\begin{abstract}

We present a study of the host galaxies of AGN selected from the
zCOSMOS survey to establish if accretion onto Supermassive Black Holes
(SMBHs) and star formation are explicitly linked up to $z\sim1$.  We
identify 152 galaxies that harbor AGN, based on their X-ray emission
($L_{0.5-10~{\rm keV}}>10^{42}$ erg s$^{-1}$) detected by $XMM-Newton$
observations of 7543 galaxies ($i_{acs}<22.5$).  Star formation rates
(SFRs), including those weighted by stellar mass, of a subsample are
determined using the [OII]$\lambda$3727 emission-line luminosity,
corrected for an AGN contribution based on the observed
[OIII]$\lambda$5007 strength or that inferred by their hard (2-10 keV)
X-ray luminosity.  We find that an overwhelming majority of AGN host
galaxies have significant levels of star formation with a distribution
spanning $\sim1-100$ M$_\sun$ yr$^{-1}$; their average SFR is higher
than that of galaxies with equivalent stellar mass ($M_*>
4\times10^{10}$ M$_{\sun}$).  The close association between AGN
activity and star formation is further substantiated by an increase in
the fraction of galaxies hosting AGN with the youthfulness of their
stars as indicated by the rest-frame color (U-V) and spectral index
$D_n(4000)$; we demonstrate that a mass-selected sample is required to
alleviate an artifical peak in the AGN fraction falling in the
transition region due to the fact that many 'blue cloud' galaxies have
low mass-to-light ratios in luminosity-limited samples.  We also find
that the SFRs of AGN hosts evolve with cosmic time in a manner that
closely mirrors the overall galaxy population and naturally explains
the low SFRs in AGNs ($z<0.3$) from the SDSS.  We conclude that the
conditions most conducive for AGN activity are a massive host galaxy
and a large reservoir of gas.  Furthermore, a direct correlation
between mass accretion rate onto SMBHs and SFR is shown to be weak
although the average ratio ($\sim10^{-2}$) is constant with redshift,
effectively shifting the evidence for a co-evolution scenario in a
statistical manner to smaller physical scales (i.e., within the same
galaxies).  The order-of-magnitude increase in this ratio compared to
the locally measured value of $M_{BH}/M_{bulge}$, is consistent with
an AGN lifetime substantially shorter than that of star formation.
Our findings illustrate an intermittent scenario with underlying
complexities regarding fueling over vastly different physical (and
temporal) scales yet to be firmly determined.

\end{abstract}





\section{Introduction}

The remarkable similarity between the evolution of the star formation
history of galaxies and the emissivity of AGN from $z\sim1$ to the
present \citep[e.g., ][]{bo98,fr99,me04,sh07,si08b} suggests a common
mechanism that regulates their growth.  Major mergers of galaxies
\citep{mi96} and secular evolution \citep{ko04,hop06} can both
potentially funnel gas to the nuclear region that can then power
concurrent star formation and accretion onto supermassive black holes.
Fully consistent with this scheme, young stellar populations are known
to be prevalent within the bulges of nearby Seyfert galaxies
\citep[e.g.,][]{te90,go01,wh03} and luminous quasars
\citep[]{ja04a,let07}.  Enhanced levels of ongoing star formation are
also evident in nearby Seyfert 2s \citep{gu06} and the hosts of some
mid-infrared selected quasars \citep{la07} although are not high
enough to put them in the same class as the Ultraluminous Infrared
Galaxies \citep{st06}.  On the other hand, the lack of star formation
based on [OII] emission in the hosts of a significant sample of PG
quasars \citep{ho05} and type 1 AGN \citep{ki06} from the Sloan
Digital Sky Survey (SDSS) may indicate an elevated role for AGN in
suppressing star formation \citep[][]{gr04,sp05,cr06,cav07,hop08a}.
However, there do appear to be high levels of star formation in the
hosts of PG quasars when considering their far-infrared emission
\citep{sc06,ne07}, and type 2 quasars from the SDSS
\citep[$0.3<z<0.8$; ][]{za03,ki06} that may hint at an underlying
evolutionary sequence (i.e., a modification to the unified model with
a correlation between star formation and nuclear obscuration) or the
systematic increase in the global star formation history of galaxies
\citep[e.g., ][]{li95,ho06,no07,tr07,zh07}.  These seemingly
discrepant results highlight the complexity in determining if a casual
relationship between star formation and AGN activity exists.  This is
most likely due to the challenges in measuring the host galaxy
properties under the glare of a bright AGN, widely varying selection
methods, and inadequate control samples.

The ability to study the host galaxies of AGN selected from a large
parent sample of galaxies has dramatically improved in the last few
years.  The SDSS has generated an unprecedented sample of galaxies to
select large numbers of low redshift ($z<0.3$) narrow-line AGN
\citep{ka03b,ke06,sta06,he06,wild07} to cleanly study their host
properties.  Powerful AGN have been clearly shown to reside in
bulge-dominated galaxies with young stellar populations \citep{ka03b}.
In conjunction with GALEX observations, AGN have optical-ultraviolet
colors placing them in an intermediate region \citep {ma07,sa07,sch07}
between the red and blue galaxy populations thus stirring interest in
their relation to transitional galaxies \citep{hop07}.  Remarkably,
\citet{ka07} find a strong correlation between the mass accretion rate
onto SMBHs and the presence of young stars in the bulge that is
typically accompanied by a star-forming outer disk that may be
supplying the nuclear fuel.  It is imperative to extend such studies
to epochs closer to the peak ($z\sim1$) of the cosmic AGN and star
formation history since AGNs may have an impact on the gas content of
galaxies and their subsequent demise.

X-ray emission is a unique probe of AGN activity due to its ability to
penetrate a moderately obscuring medium ($N_H \lesssim 10^{24}$
cm$^{-2}$), the lack of confusion with stellar sources, and because it
allows a more direct probe of the mass accretion rate \citep[see][for
a review]{br05}.  Selection of narrow line AGN (i.e., obscured; type
2) by X-rays \citep[e.g., ][]{ba03,sz04,si05,br07} can compensate for
the deficiencies of optical selection based on emission line ratios at
$z>0.3$.  X-ray observations with both $Chandra$ and XMM-$Newton$ are
being exploited to cleanly study the host galaxies of X-ray selected
AGN \citep{gr05,na07,pi07,ga08}.  Recently, \citet{si08a} show that the
bulge-dominated hosts of X-ray selected AGN in the Extended $Chandra$
Deep Field - South have rest-frame colors that are bluer with
increasing redshift, possibly related to the star formation history of
galaxies.  In a related study, the mean star formation rate based on
broad-band photometry of 58 X-ray selected AGN ($z\sim0.5-1.4$) in the
1 Msec CDF-S has been shown to be similar to IRAC-selected galaxies in
an equivalent mass regime \citep{ah07}.  To date, few studies using a
significant sample of AGN selected from a large parent sample of
galaxies with optical spectroscopy have been undertaken at these
higher redshifts ($z>0.3$).

To do so, we utilize the rich multiwavelength observations of the
COSMOS field \citep{sc07} to carry out such a study at $0.5 \lesssim z
\lesssim1.0$.  The COSMOS survey is roughly a 2 square degree region
of the sky selected to be accessible from all major observatories both
from the ground (e.g., Subaru, VLT) and space (e.g., $HST$, $Spitzer$,
XMM-$Newton$, $Chandra$).  The zCOSMOS survey \citep{lilly07} targets
objects for optical spectroscopy with the VLT in two separate
observing programs. A bright sample ($i<22.5$) is observed with a red
grism to provide a wavelength coverage of $5500-9500$ {\rm \AA}
ideal for identifying galaxies ($L_*$) up to $z \sim 1.2$.  A deep
program, not utilized in the present study, targets faint galaxies
($B<25$), selected to be in the redshift range $1.5\lesssim z\lesssim
2.5$, using a blue grism with a wavelength coverage of
$3600<\lambda<6700$ {\rm \AA}.  Here, we select a sample of galaxies
based on their stellar mass with reliable spectroscopic redshifts from
the zCOSMOS bright program.  Those that host AGN are identified by
their X-ray emission as detected by XMM-$Newton$ \citep{ha07,cap07}.
We measure the strength of emission lines (i.e., [OII]$\lambda$3727,
[OIII]$\lambda$5007), using our automated pipeline
\citep["platefit\_vimos";][]{lam08} to determine star formation rates
for the entire galaxy sample including those hosting AGN.  An
additional spectral indicator ($D_n4000$) enables us to discern the
age of the stellar populations on longer timescales.  X-ray
luminosities of the AGN give us a handle on their bolometric output,
less affected by obscuration, to infer mass accretion rates and
determine any trends with star formation rate and redshift.  Finally,
we point out that a companion paper \citep{si08c} based on the zCOSMOS
10k catalog expands on the current study by investigating the
environmental impact on AGN activity and their host galaxy properties.

Throughout this work, we assume $H_0=70$ km s$^{-1}$ Mpc$^{-1}$,
$\Omega_{\Lambda}=0.75$, and $\Omega_{\rm{M}}=0.25$.

\section{Data and derived broad-band properties}

\subsection{Parent galaxy sample and AGN identification}

We use the zCOSMOS 10k spectroscopic 'bright' catalog to construct a
well-defined sample of galaxies including those hosting X-ray selected
AGN up to $z\sim1$.  Specifically, we identify 7543 galaxies with a
selection magnitude $i_{ACS} \leq 22.5$ and high quality spectra hence
reliable redshifts up to $z=1.02$.  A galaxy is included in our sample
if the redshift has a quality flag 2.0 or higher that amounts to a
confidence of $\sim99\%$ for the overall sample.  The redshift success
rate is not strongly dependent on galaxy color over $0.5\lesssim z
\lesssim1.0$ given the quality of the spectra and presence of strong
features (i.e., 4000 \AA~break; Ca H+K, [OII]).  Full details on data
acquisition, reduction, redshift measurements and quality assurance
can be found in \citet{lilly07,lilly08}.

We use the catalog of X-ray sources \citep{cap07} generated from the
uniform XMM-$Newton$ coverage ($\sim50$ ks depth) of the COSMOS field
\citep{ha07} to identify galaxies within our parent sample that harbor
AGN.  The X-ray catalog includes 1848 point--like sources detected
above a given threshold using a maximum likelihood method in either
the soft (0.5--2 keV), hard (2--10 keV) or ultra-hard (5--10 keV)
bands down to a limiting flux of 5$\times 10^{-16}$, 2$\times
10^{-15}$ and 5$\times 10^{-15}$ erg cm$^{-2}$ s$^{-1}$ in the
respective band.  The adopted threshold ($Likelihood > 10$)
corresponds to a probability $\sim 4.5\times10^{-5}$ that a source is
a background fluctuation.  An additional 26 faint XMM sources
coincident with diffuse emission (Finoguenov et al. in preparation)
were excluded.  Both the soft (0.5-2.0 keV) and hard band (2.0-10.0
keV) detections comprise our AGN sample in order to include the
low-to-moderate luminosity and/or absorbed sources.  The higher
sensitivity of the XMM-$Newton$ observations in the soft band enables
us to probe lower luminosity AGN ($log~L_X\sim43$) with $0.8\lesssim z
\lesssim 1.0$ not sufficiently represented in the hard-band catalog.

Optical and near-infrared counterparts to the XMM-$Newton$ X-ray
sources are found using a maximum-likelihood method \citep{br07}.
Most X-ray sources (84\%) have reliable counterparts while the
remaining are uncertain due to the presence of multiple objects within
the X-ray error box that have similar probababilities of being the
true counterpart.  We use the $Chandra$ observations (Elvis et al. in
preparation; Civano et al. in preparation) that cover the central 1
square degree to further refine our identifications.  We elect to
include counterparts with lower confidence given that $\sim50\%$ of
them are likely to be correct based on the overlap between
XMM-$Newton$ and $Chandra$ identifications (M. Brusa, private
communication).  All results presented here are confirmed based on the
slightly smaller catalog of highly reliable counterparts.  

The zCOSMOS 'Bright' program provides optical spectra for 357 of the
1093 optical counterparts to X-ray sources having $i<22.5$.  Reliable
spectroscopic redshifts are available for 90\% of which 164 have
redshifts between $0<z<1.02$.  Higher redshift AGN up to $z\sim4$ have
been identified by the zCOSMOS 'Bright' program but are not the focus
of this study.  Since a high fraction of the zCOSMOS galaxies with
associated X-ray emission have detections in the hard 2-10 keV band
(84\%), we do not expect the inclusion of soft-selected sources to
induce any significant bias.  A subset (54\%) of the X-ray sources has
been designated as 'Compulsory' targets when designing the VIMOS masks
which essentially lessens their randomness and results in a 71.9\%
sampling rate compared to the current rate of 29.8\% for galaxies.  We
set priority to these sources since they were not included in the
spectroscopic followup program using Magellan \citep{trump07}.  Since
the number of AGNs is not overwhelmingly large, we chose to not select
a smaller pseudo-random sample but rather incorporate correction
factors applied to derived measurements where applicable.

\begin{figure}

\epsscale{1.2}

\plotone{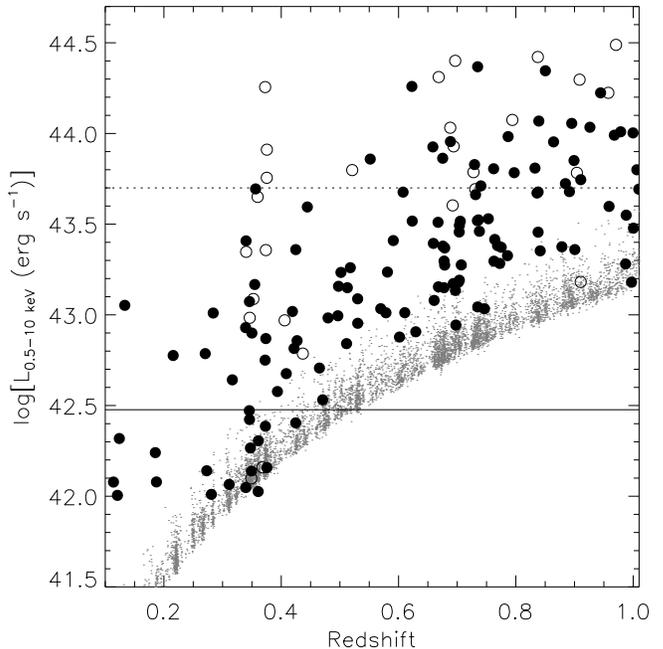}
\caption{X-ray luminosity (0.5-10 keV) versus redshift for 152 AGN
(large symbols).  Filled circles are AGN primarily characterized by
narrow emission lines while open symbols denote those with at least one
broad optical emission line falling within the observed spectral
range.  The upper limit to the broad-band X-ray luminosity for each
zCOSMOS galaxy is shown by the small grey dots.  The solid horizontal
line is the minimum X-ray luminosity that we enforce to measure the
fraction of galaxies hosting AGN to ensure a significant underlying
galaxy population.  The dotted line shows our chosen upper limit to
avoid AGN contamination.}

\label{lx_z}
\end{figure}

Moderate-luminosity AGN ($L_X\sim10^{43}$ erg s$^{-1}$), detected in
medium-to-deep X-ray observations, have been shown to be ideal
laboratories for the study of their host galaxies due to their low
optical brightness which in many cases is due to their nuclear
obscuration \citep[e.g.,][]{to06,main07}.  Our AGNs predominately have
luminosities below that of typical QSOs ($L_X>10^{44}$ erg s$^{-1}$)
but are the dominant contributor to the Cosmic X-ray Background
\citep{gi07}.  In specific cases, we further select AGN based on their
X-ray luminosity ($42.0 < log~L_{0.5-10.0~{\rm keV}}<43.7$), as done
in \citet{si08a}, to isolate a sample for which we can cleanly
determine their host properties (i.e., stellar masses, rest-frame
$U-V$, $D_n4000$).  \citet{ga08} have demonstrated that the optical
emission from a nearly equivalent AGN sample in COSMOS is primarily
attributed to their host galaxies; AGN dominated galaxies are
preferentially at $z>0.8$ due to the fact that their X-ray
luminosities are typically above our cutoff ($L_X\sim10^{43.7}$ erg
s$^{-1}$).  We conclude that the derived properties of the host
galaxies of AGN in zCOSMOS are likely to be reliable.  Where feasible,
we include higher luminosity AGN (QSOs; 16 with $L_{0.5-10~{\rm
keV}}>10^{44}$ erg s$^{-1}$), since we can measure their stellar
properties (i.e., [OII] strength) even under the glare of an
intrinsically, bright AGN (see Section~\ref{sfr_measure} for details).
For these cases, we refrain from analyses based on their host
properties such as mass-weighted SFR or rest-frame color.  One caveat
of our luminosity-selected sample is a potential bias towards specific
accretion modes such as the "Seyfert mode" \citep{hop06} or specific
evolutionary stages in their fueling such as pre- or post-mergers.
Fortunately, we can test the later since merger-driven models
\citep{hop08a} predict vastly different properties (i.e., SFRs,
morphology) of the hosts of AGNs before and after the coalescence of
massive disk galaxies.

We list in Table~\ref{sample} the statistics of the parent galaxy
sample and those hosting X-ray selected AGN.  The redshift and X-ray
luminosity distribution of the full AGN sample (152) is shown in
Figure~\ref{lx_z} up to $z=1.02$.  Symbols denote their optical
spectral properties with 82\% lacking broad ($FWHM>1000$ km s$^{-1}$)
emission lines and most characterized as narrow emission-line
galaxies.  As further detailed in subsequent sections, we isolate
various subsamples based on redshift, galaxy mass, and AGN luminosity
to optimize our analyses.  For instance, as described above, we
enforce an upper limit to the X-ray luminosity
($log~L^{max}_{X}=43.7$) of the AGN when measuring a quantity with
respect to its host galaxy.  Also, we slightly increase our minimum
X-ray luminosity ($log~L^{min}_{X}$=42.48) when determining the
fraction of galaxies hosting AGN to ensure that a significant parent
sample of zCOSMOS galaxies are capable of detecting each AGN to avoid
effects based on limited statistics\footnote{Specifically, an
overabundance of AGN at $z\sim0.35$ falling at the flux limit of the
$XMM$-Newton observations, seen in Figure~\ref{lx_z}, have an adverse
effect on our determination of the fraction of galaxies hosting an
AGN, due to the limited sample of zCOSMOS galaxies capable of
detecting AGN with $L_X\sim10^{42}$ erg s$^{-1}$, that disappears when
implementing a slightly higher selection on luminosity.}.

\subsection{Stellar mass measurements}

\begin{figure}
\epsscale{1.2}

\plotone{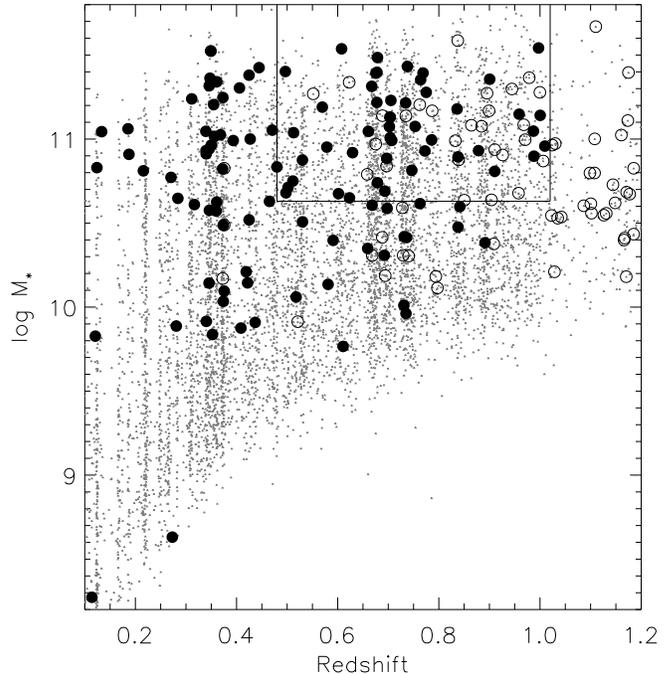}
\caption{Stellar-mass versus redshift for 7543 zCOSMOS galaxies (small
grey circles). The box marks the mass-selected subsample of galaxies
($log~M>10.6$; $0.48<z<1.02$) for which (specific) star formation
rates are determined using [OII]$\lambda$3727.  Galaxies hosting X-ray
selected AGN are further marked by a larger black circle (filled:
$42.0<log~L_{0.5-10~{\rm keV}}<43.7$; open: $log~L_{0.5-10~{\rm
keV}}>43.7$).}

\label{selection}
\end{figure}

\begin{figure*}
\includegraphics[angle=90,scale=0.75]{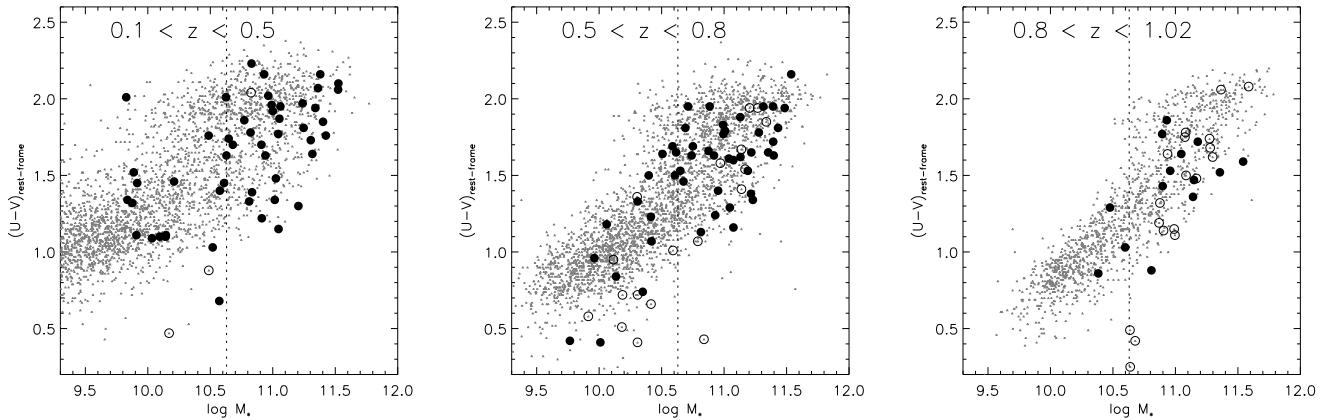}

\caption{Rest-frame color $U-V$ versus stellar mass split into three
redshift intervals for galaxies and AGN with symbols described in
Figure~\ref{selection}.  The vertical line marks our imposed mass
limit.}

\label{color_mass}
\end{figure*}

Stellar masses, including rest-frame absolute magnitudes (AB system;
$M_U$, $M_V$), are derived from fitting stellar population synthesis
models from the library of \citet{bc03} to both the broad-band optical
\citep[CFHT: $u$, $i$, $K_s$; Subaru: $B$, $V$, $g$, $r$, $i$,
$z$;][]{capak07} and near-infrared \citep[$Spitzer$/IRAC: $3.6 \mu $,
$4.5 \mu $;][]{sand07} photometry using a chi-square minimization for
each galaxy.  The measurement of stellar mass (M$_{*}$) includes (1)
the assumption of a Chabrier initial mass function, (2) a star
formation history with both a constant rate and an additional
exponentially declining component covering a range of time scales
($0.1 < \tau < 30$ Gyr), (3) extinction ($0<A_V<3$) following
\citet{ca00}, and (4) solar metallicities.  Further details on mass
measurements can be found in \citet{bo08} and \citet{me08}.  For
compatibility with the star formation rate calibration of \citet{mo06}
implemented in subsequent sections, we convert all masses to an
equivalent based upon a Salpeter IMF by applying a multiplicative
factor of 1.7 \citep{po07}.  We show the mass distribution for the
parent galaxy population, including those harboring X-ray selected
AGN\footnote{It is worth mentioning that there may be a potential
problem that galaxies with even moderate-luminosity AGN may have
inaccurate mass estimates; a bluer continuum will essentially reduce
the stellar age and hence lower the mass measurements since the
derived mass-to-light ratio depends strongly on the spectrum.}  with
$log~L_{0.5-8.0~{\rm keV}}>42.0$, as both a function of redshift
(Figure~\ref{selection}) and rest-frame color ($U-V$;
Figure~\ref{color_mass}).  All masses are expressed in solar units
(M$_{\sun}$) throughout this work.

We determine a minimum mass threshold that all galaxies must satisfy
up to $z\sim 1.0$.  The mass limit is set in order to ensure a fairly
complete representation of both blue and red galaxies at all redshifts
considered.  In Figure~\ref{color_mass}, it is clearly evident that
the mass limit of $log~M=10.6$ is essentially imposed by the red
galaxy population at $z\gtrsim0.8$ (rightmost panel).  The lack of red
galaxies below this limit is due to our initial selection on apparent
magnitude.  \citet{me08} estimate based on a series of mock catalogs
from the Millennium simulation that the zCOSMOS 'Bright sample' is
essentially complete for galaxies with $log~M\approx10.6$ at $z=0.8$
while the completeness drops to $\sim50\%$ at $z=1$.  In total, we
have a sample of 2540 galaxies ($0.1<z<1.02$) above this mass limit of
which 105 host X-ray selected AGN (see Table~\ref{sample} for
statistics regarding subsamples employed herein.).

\section{[OII] as a star-formation rate indicator and the AGN contribution}
\label{sfr_measure}

Our primary aim is to use the emission line [OII]$\lambda$3727 to
measure the SFRs of a well-defined sample of galaxies including those
hosting AGN, as commonly used for star-forming galaxies at these
redshifts \citep[e.g., ][]{ke04,co07}.  The availability of [OII]
within the spectral window of the zCOSMOS 'bright' program requires
our sample of galaxies to fall within redshift range $z=0.48-1.02$.
The [OII] line is one of a suite of spectral features (e.g.,
H$\alpha$$\lambda$6563, H$\beta$$\lambda$4861, [OIII]$\lambda$5007)
measured by an automated routine "platefit\_vimos" \citep{lam08} that
simultaneously fits all lines with a gaussian function to determine
line flux and equivalent widths for the entire zCOSMOS 10k sample.
Emission line fluxes are corrected for slit loss based on a comparison
of their spectroscopic and photometric ($i_{ACS}$) magnitudes.  Line
measurements with a significance less than 1.15$\sigma$ are quoted
here as upper limits.  

An assessment of the AGN contribution to the [OII] emission line is
required even though the production of this low ionization line has
been shown to be relatively weak. \citet{cr02} postulate, based on a
comparison of the equivalent width of [OII] in the composite 2dF QSO
spectrum to that in a composite 'normal' galaxy spectrum, that [OII]
is mainly produced by star-formation in the host galaxies of AGN over
a wide range of absolute magnitude.  The [OII] strength has been
observed to be $\sim10-30\%$ of the [OIII]$\lambda$5007 emission line
flux \citep{fe86,ho05,ki06}.  \citet{ho05} has explored the
feasibility of using [OII] as an indicator of ongoing star formation
for a sample of PG quasars that in many cases provided only an upper
limit.  Upon further investigation using emission line ratio
diagnostics, \citet{ki06} conclude based on a sample of $\sim3600$
type 1 AGN ($z<0.3$) selected from the SDSS that [OII] emission is
mainly attributed to AGN photoionization and does not have a strong
HII component.    

\begin{figure*} 

\includegraphics[angle=90,scale=0.75]{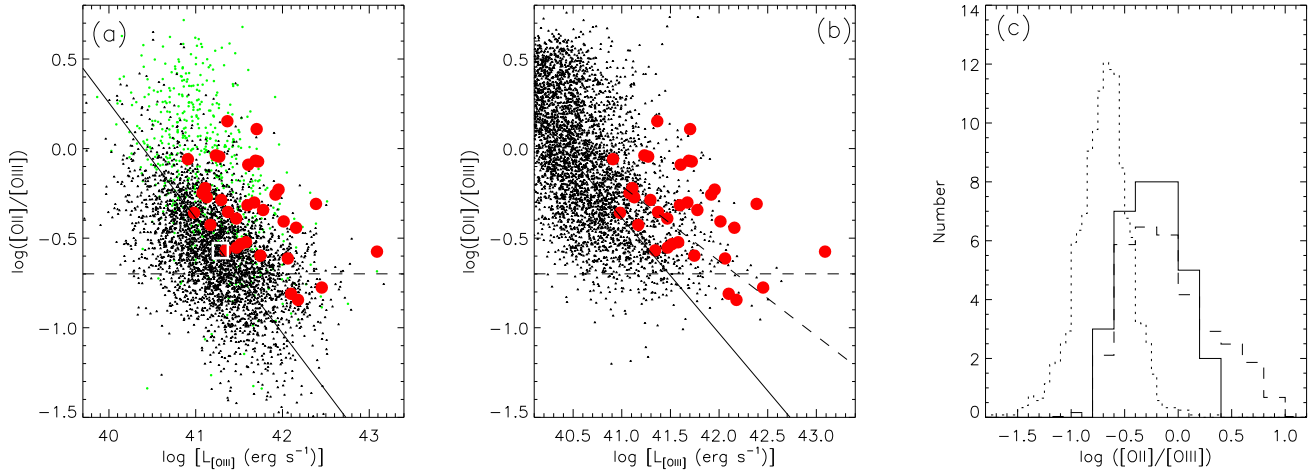} 

\caption{Emission-line properties of AGN.  ($a$) Observed (i.e., no
extinction correction) line ratio [OII]/[OIII] versus [OIII]
luminosity for zCOSMOS galaxies ($log~M>10.6$; small green points) and
those hosting AGN shown by larger red circles.  For comparison,
emission line properties of type 1 AGN from the SDSS \citep[$z<0.3$;
][]{ki06} are marked by the small black dots that are further
characterized by the best fit linear relation (solid line) and mean
ratio $<[OII/[OIII]>=0.27$ (open white square).  The dashed horizontal
line denotes our single assumption for the value of a purely AGN
dominated [OII]/[OIII] ratio (see text for further details).  ($b$)
Same as panel $a$ but with type 1 AGNs replaced by type 2 AGNs from
\citet{ka03b}.  The slanted lines are the best fit linear relation to
type 1 (solid) and type 2 (dashed) AGNs.  ($c$) Histogram of the
extinction-corrected [OII]/[OIII] distribution for the type 1/2 AGN
(short dash/long dash) from the SDSS and our zCOSMOS sample (solid)
all having $log~L_{\rm [OIII]}>41.3$.}

\label{emlines} 
\end{figure*}

The aforementioned results do not rule out the potential effectiveness
of using [OII] as a SFR indicator for different samples of AGNs such
those having higher luminosities, significant intrinsic obscuration or
being at higher redshifts.  For example, the stellar populations of
'strong' type 2 AGN from the SDSS have young ages similar to
late-type galaxies \citep{ka03b} based on the age indicator $D_n4000$.
We illustrate that this is likely related to the presence of ongoing
star formation.  In Figure~\ref{emlines}, we show the [OII]/[OIII]
ratio for type 1 (panel $a$) and type 2\footnote{The luminosity
selection of type 2 AGNs from the SDSS is equivalent to the
definition of 'strong' AGN ($L>10^7$ L$_{\sun}$) given in
\citet{ka03b}.  This luminosity selection, based on
extinction-corrected values, guarantees that the sample is dominated
by Seyfert galaxies comparable to the luminosities of the zCOSMOS
AGN.}  (panel $b$;$L_{\rm [OIII]}>10^{40.58}$ erg s$^{-1}$) AGNs from
the SDSS having $z<0.3$.  The data for 25000 type 2 AGNs is taken
from the high level products available from MPA based on the SDSS DR4
release; we require a $S/N>3$ detection for [OII], H$\beta$, [OIII]
and H$\alpha$.  First of all, it is worth highlighting that the
best-fit linear relation to the type 1 AGN (panel $a$) illustrates a
scenario where [OII] emission scales with [OIII] ($log~L_{\rm
[OII]}\propto 0.36\times L_{\rm [OIII]}$) but not with a one-to-one
correspondence.  Even without correcting for extinction, we see that
there is evidence (Fig.~\ref{emlines}$b$) for elevated [OII]/[OIII]
for type 2 AGNs as seen by the flatter slope of the best-fit relation
(slanted dashed line; $log~L_{\rm [OII]}\propto 0.58\times L_{\rm
[OIII]}$) compared to that of the type 1 AGNs.  The change in slope
appears to signify that the strength of an additional component to
[OII] (i.e., star formation) increases with AGN luminosity possibly
related to the results of \citet{ka07} where a decline in stellar age
is seen for AGNs with higher mass accretion rates.  Finally, we see
conclusively that the AGN host galaxies in zCOSMOS, while having
higher [OIII] luminosities, have elevated [OII]/[OIII] ratios compared
to type 1 AGNs (panel $a$), and exhibit a similar slope to the type 2
AGNs in the SDSS and slightly further enhanced [OII] emission.

\begin{figure}
\epsscale{1.0}
\plotone{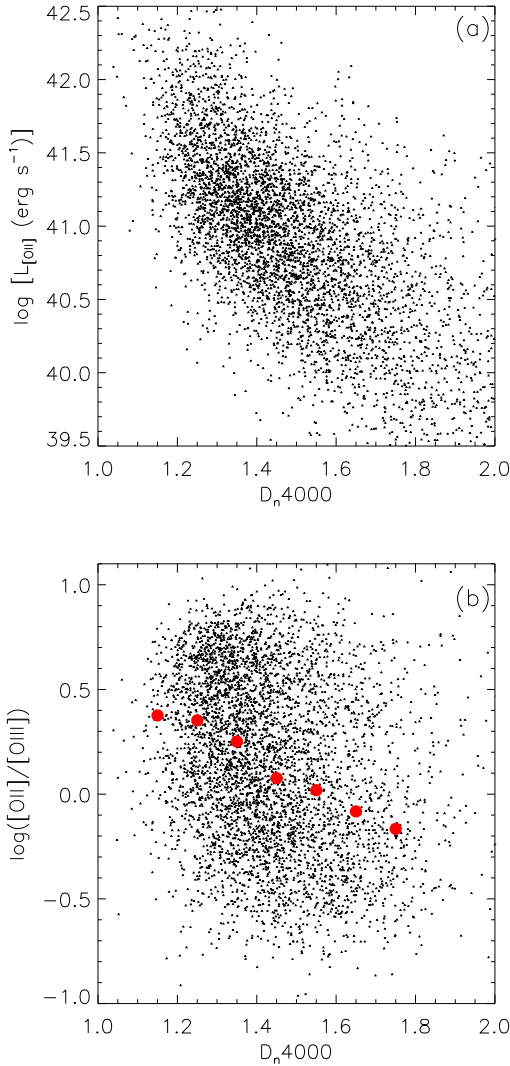}

\caption{Relation between $D_n4000$ and (a) [OII] luminosity and (b)
[OII]/[OIII] ratio for 4609 luminous($L_{\rm [OIII}> 10^7$
L$_{\sun}$), type 2 AGN from the SDSS.  In panel $b$, the median
[OII]/[OIII] ratio is shown (large red circles) in bins of $D_n4000$
having a width of 0.1.}
\label{type2_sdss}
\end{figure}

When correcting for internal extinction based on the Balmer
decrements, we find that the type 2 AGNs have a significantly higher
distribution of [OII]/[OIII] than the type 1s (Fig.~\ref{emlines}$c$).
This further suggests that star formation is more prevalant in the
type 2 AGN population compared to those with an observable broad line
region; such a hypothesis has been put forth by \citet{ki06} to
explain the [OII] emission ($<[OII]/[OIII]>=-0.12$) evident in the
more luminous, type 2 QSOs \citep{za03}.  To further justify our use
of [OII], we demonstrate in Figure~\ref{type2_sdss} that a correlation
between total [OII] emission (HII+AGN) and the spectral index
$D_n4000$ exists and is in agreement with significant levels of
ongoing star formation seen in the hosts of type 2 AGN \citep{gu06,yan06}.

For our purpose, we use the higher ionization line [OIII]$\lambda5007$
when present in our spectra, and an empirical relation between [OII]
and [OIII] to statistically remove the component of the detected [OII]
line that can be attributed to the AGN.  This entails an assumption
that the [OIII] line is purely of AGN origin \citep{ka03b} that may
lead us to underestimate SFRs due to a significant stellar
contribution but only in very few cases since the [OIII] luminosities
of Seyfert galaxies in SDSS, and also zCOSMOS AGN hosts (see
Figure~\ref{emlines}$a$), are substantially higher than that of
typical HII galaxies \citep{ke06}.  \citet{ki06} find that the median
value of the observed [OII]/[OIII] ratio is 0.27 with significant
dispersion (see Fig.~\ref{emlines}$a$).  We chose to use a slightly
lower ratio ([OII]/[OIII]=0.21) that is the mean value for type 1 AGN
in the SDSS sample (M. Kim private communication) with
$log~L_{[OIII]}> 41.5$, a luminosity regime similar to our zCOSMOS AGN
sample.  This higher luminosity cut guarantees that the ratio best
reflects that produced by the AGN, while effectively minimizing a
contribution from HII regions.  This results in slightly higher SFRs
for our AGN sample than if we chose to use the mean of the entire type
1 SDSS sample but our final results in this study are consistent using
either value.  We make no attempt to use a luminosity-dependent
correction, as inferred by the SDSS type 1 AGN, since a simple linear
relation may not be evident at higher luminosities
($log~L_{[OIII]}>42$).  Also, there is no need to correct this
relation for extinction given the low levels of dust attenuation
\citep[$A_V\approx0.2$ mag;][]{ki06} observed in these type 1 AGN.  We
strongly note, given the high dispersion in the [OII]/[OIII] ratio for
type 1 SDSS AGN, that this is only a statistical correction applied to
the population as a whole to infer the global or mean properties of
the sample; the individual measurements of [OII] strength associated
with star formation in a particular galaxy are expected to be
inaccurate.  We further note that even with this correction the more
luminous AGN tend to have higher SFRs as evident by the best-fit
relation for type 2 AGNs in the SDSS (Fig~\ref{emlines}$b$;
$log~L_{\rm [OII]}\propto 0.37\times L_{\rm [OIII]}$).  In addition,
we assess the impact of the dispersion in the [OII]/[OIII] ratio for
type 1 SDSS AGNs by assuming a normal distribution of the line ratio
($<log [OII]/[OIII]>=-0.69$; $\sigma=0.25$) and perform many
iterations in our determination the distribution of SFR.

We consider extinction due to dust as an important factor in
determining the component of the observed [OII] emission attributed to
AGN photoionization given that our sample is predominantly composed of
type 2 AGN (see Fig.~\ref{lx_z}).  An extinction-corrected [OIII]
luminosity is found for each zCOSMOS AGN with detected [OIII] line
emission using $A_V=0.8$.  Since most of our sample does not have
H$\alpha$ and H$\beta$ within our observed spectral window, we chose
to implement a level of extinction\footnote{This amount of optical
attenuation is similar to that of X-ray selected AGN; we find a median
$A_V$ of 0.98 based on the Balmer decrements of 20 AGN in our sample
with significant $H\alpha$ and $H\beta$ line measurements (Mainieri et
al. in preparation).} based on the mean Balmer decrement of type 2
Seyferts in SDSS \citep{ke06}.  This same level of attenuation is then
reapplied to the inferred AGN component to [OII], based on the
empirical relation mentioned above, and subtracted from the observed
(i.e. no dust correction) [OII] emission line luminosity to provide an
estimate free of any AGN contribution.  We are confident that this
method is applicable to the zCOSMOS sample given that the [OII]
luminosities of AGN hosts are all systematically higher
($\sim3-4\times$) than that expected from gas photoionized by an AGN
as indicated by the SDSS type 1 AGN (Fig.~\ref{emlines}$c$) of similar
luminosities ($log~L_{\rm [OIII]}>41.3$).  Also, we highlight that the
enhancement of [OII] relative to [OIII] for our zCOSMOS AGN is not
purely induced by our extinction corrections as evident in observed
relation (i.e., no extinction correction; Fig.~\ref{emlines}$a,b$).

\begin{figure} 
\epsscale{1.15}
\plotone{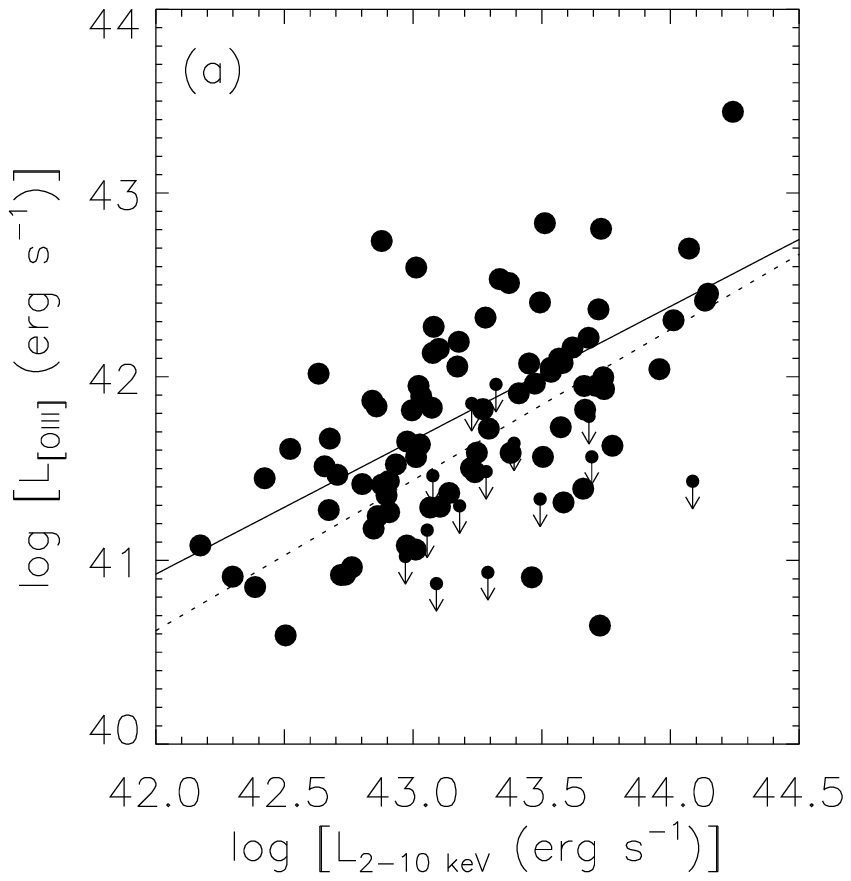}
\epsscale{1.05}
\plotone{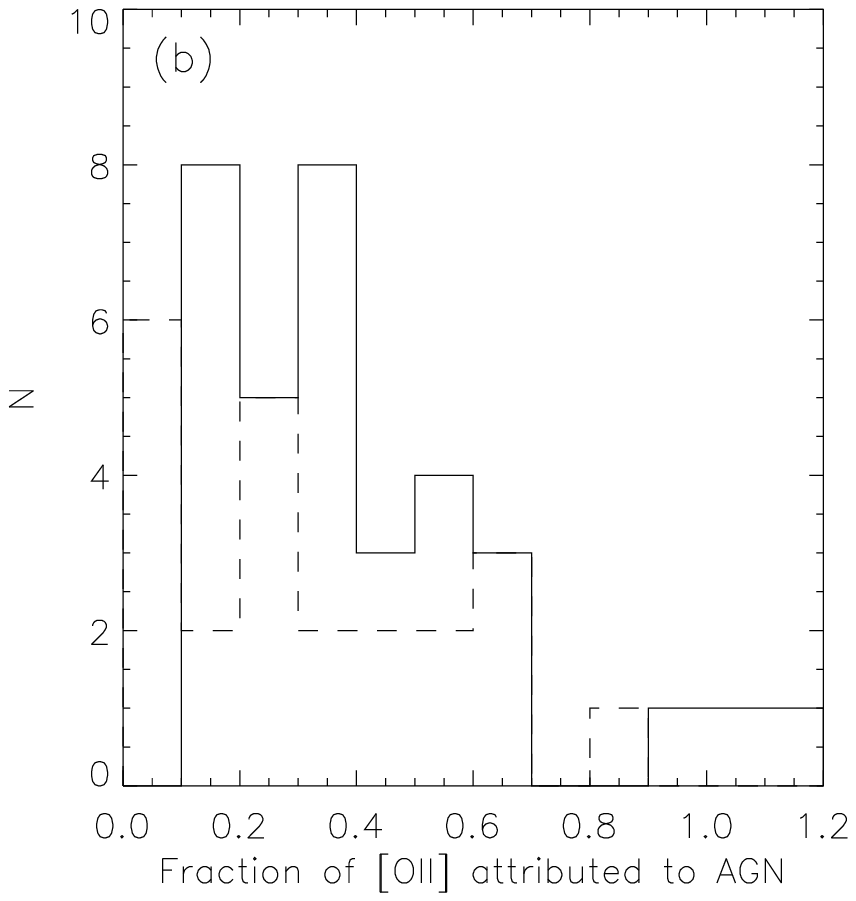}

\caption{($a$) Extinction-corrected [OIII]$\lambda$5007 luminosity
versus X-ray luminosity (2-10 keV) for AGN.  A clear correlation is
evident as shown by the best-fit linear relation to these data (solid
line) and local Seyferts \citep[dotted line; ][]{pa06}.  ($b$)
Fraction of the [OII] line luminosity attributed to an AGN and split
by the method of determining the AGN contribution.  The number
distribution of galaxies with detected [OIII] emission, used to
estimate the amount of [OII] due to AGN activity, is given by the
solid histogram.  Galaxies without [OIII] detections are indirectly
assessed by their hard X-ray luminosity (dashed histogram).  There is
no evidence for a systematic offset between both methods.}

\label{o3_lx} 
\end{figure}

For AGN host galaxies at $z\gtrsim0.8$, we need to modify our method
since [OIII] is no longer within our observed spectral window.
Fortunately, we can utilize the strong correlation between hard X-ray
(2-10 keV) and [OIII] luminosity \citep{he05,pa06} to estimate the AGN
contribution for these cases.  It is important to highlight that our
AGN are X-ray selected by the 0.5-10 keV band thus limiting the
inclusion of more heavily absorbed type 2 AGN such as the
Compton-thick population \citep{fi08}, for which this correlation is
less evident if the X-ray luminosities are not corrected for X-ray
absorption \citep{he05} as is the case here.  In
Figure~\ref{o3_lx}$a$, we plot the [OIII] line luminosity, corrected
for extinction as done in the previous paragraph, as a function of the
rest-frame 2-10 keV luminosity for AGN in the zCOSMOS sample with
redshifts over a larger baseline ($0.2<z<1.0$).  We use these lower
redshift AGN in order to provide higher statistics to measure the
best-fit linear relation between these two quantities.  We determine
the best fit $L_X-L_{[OIII]}$ relation (Equation~\ref{lx-o3}) for our
sample by implementing a bivariate linear regression (EM) algorithm
with ASURV \citep[Survival Analysis for Astronomy package Rev. 1.2;
][]{la92} that considers limits as well as detections.

\begin{equation}
log~L_{[OIII]}=(0.729\pm0.101)~log~L_{2-10~keV}+(10.307\pm4.381)
\label{lx-o3}
\end{equation}

\noindent This relation (Fig.~\ref{o3_lx}$a$), as shown by the solid
line, is similar to previous studies based on a sample of low redshift
AGN \citep[e.g.,][]{pa06} shown by the dotted line.  We find a nearly
equivalent slope although a higher normalization (i.e., larger values
of $L_{{\rm [OIII]}}$ for a given $L_{2-10~{\rm keV}}$).  The strong
linear relation (Pearson correlation coefficient $r=0.63$) strengthens
our assumption that most of the [OIII] line luminosity is due to AGN
photoionization.  Any significant contribution from stars to the
[OIII] luminosity would increase our derived star formation rates in
AGN hosts.  In Figure~\ref{o3_lx}$b$, we demonstrate that there are no
systematic offsets by introducing this indirect probe of the [OIII]
line luminosity; both methods essentially remove a similar fraction
($\lesssim 60\%$) of the [OII] luminosity attributed to AGN
photoionization.

We determine SFRs using the empirical calibration of \citet{mo06}.
This relation considers interdependent factors (i.e., dust extinction,
metallicity, and ionization) to provide a SFR tracer that is
applicable to samples for which these quantities are inadequately
known.  To derive the following empirical relation, we fit a linear
relation to the bright end ($log~L(B)>9.5$) of the dependence of SFR
on absolute $B$-band magnitude \citep[$M_B$; see Figure 19 of
][]{mo06}:

\begin{equation}
log~SFR({\rm [OII]})=log~L_{{\rm [OII]}}-41-0.195 \times M_{B}-3.434
\label{eq:sfr}
\end{equation}

\noindent The AGN component to the emission line luminosity
$L_{[OII]}$ is removed as described above and no correction for dust
extinction is applied that is in essence an assumption that star
formation is external to a dusty NLR and circumvents the potential
problem that extinction of the nuclear region may differ from that of
HII regions \citep[see][]{gu06}.  We do not attempt to adjust (i.e.,
lower) the upper limits on nondetections by considering the
contribution of an AGN.  It is worth noting that an additional caveat
of our method is that there is an underlying assumption that the
metallicity and extinction of galaxies with or without AGN are
similar.

\section{Physical properties of AGN host galaxies}

\subsection{Stellar masses}
\label{mass}

\begin{figure}
\epsscale{1.1}
\plotone{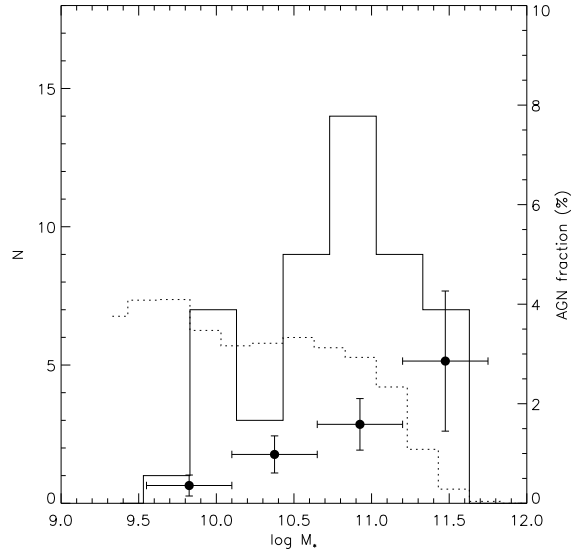}
\caption{Stellar mass distribution of galaxies hosting AGN in the
redshift range $0.1<z<0.5$.  Fifty-two galaxies ($log~M_*>9.5$) with
X-ray selected AGN ($42<Log~L_{0.5-10~{\rm keV}}<43.7$) are shown by
the solid histogram.  The observed parent distribution of 3356
galaxies (dotted histogram) is renormalized to match the AGN
distribution.  The data points represent the AGN fraction
($42.48<Log~L_{0.5-10~{\rm keV}}<43.7$; scale on the right hand
vertical axis) in fixed mass intervals (shown by the horizontal bars
centered on each data point) with $1\sigma$ error bars.}

\label{mass_frac}
\end{figure}

We measure the fraction of galaxies that host X-ray selected AGN as a
function of their stellar mass to determine the characteristic mass
above which AGN activity is most prevalent.  First of all, it is
apparent in Figure~\ref{selection} that galaxies hosting X-ray
selected AGN are preferentially massive ($log~M\gtrsim10.5$) as
evident over the full redshift range and in agreement with most
studies to date \citep[e.g.,][]{ka03b,bu08}.  For this exercise, we
consider the redshift interval $0.1<z<0.5$ for which we have a high
degree of completeness for galaxies over a wide mass baseline ($log~M
\gtrsim 9.5$).  We follow the technique discussed in $\S$~3.1 of
\citet{le07} to determine the AGN fraction of our parent population of
galaxies that accounts for the spatially varying sensitivity limits of
the {\it XMM} observations of the COSMOS field \citep[see Fig.~17
of][]{cap07}.  The necessity of this approach is demonstrated in
Figure~\ref{lx_z}, which shows the limiting X-ray luminosity as a
function of redshift for the entire galaxy sample.  Even though the
sensitivity of the $XMM$ coverage is remarkably uniform, some
dispersion is present as shown by the relatively narrow distribution
of the X-ray upper limits at each redshift.  To properly account for
the luminosity-redshift relation, we determine the contribution of
each AGN separately to the total fraction.  The AGN fraction ($f$; see
equation{~\ref{fraction} below) is determined by summing over the full
sample of AGN ($N$) with $N_{\rm gal,i}$ representing the number of
galaxies in which we could have detected an AGN with X-ray luminosity
$L^{i}_{\rm X}$.  The different sampling rates, based soley on slit
placement, of the galaxies ($S_{gal}$) and AGN (i.e., X-ray
sources\footnote{In many cases, X-ray sources have redshifts based on
a random placement of a slit for which the sampling reflects the
overall galaxy population ($S_x=S_{gal}$)}, $S_x$) are taken into
account in equation~\ref{fraction}.  We estimate the associated
1$\sigma$ error (equation~\ref{fraction_error}) using binomial
statistics where $N^{eff}_{agn}$ is the number of AGN that would be
detected if all galaxies have the same limiting X-ray sensitivity and
the sample of AGN was randomly selected.  Here, we only consider AGN
with $42.48<log~L_{0.5-10 {\rm keV}}<43.7$.  As previously mentioned,
the lower limit ensures that we have a statistically significant
sample of parent galaxies ($\gtrsim700$) that could host each AGN
while the upper limit restricts the sample to low-to-moderate
luminosities thus securing the accuracy of their host-galaxy masses.
We refer the reader to \citet{si08a} for further details and results
employing this method.

\begin{equation}
f=\sum_{i=1}^{N}\frac{1/S_x}{N_{\rm gal,i}/S_{gal}}
\label{fraction}
\end{equation}

\begin{equation}
\sigma^2=N^{eff}_{agn}\times(N_{gal}-N^{eff}_{agn})/N^3_{gal}
\label{fraction_error}
\end{equation}

In Figure~\ref{mass_frac}, we plot the fraction of galaxies that host
AGN and the number distribution (Table~\ref{sample}; Sample A) of both
galaxies and those with AGN over a slightly wider luminosity interval
($42.0<log~L_{0.5-10 {\rm keV}}<43.7$).  The absolute fraction is low
($\sim1-4\%$) due to our restrictions on X-ray luminosity.
\citet{ka03b} measure a fraction $\sim5\times$ higher although they
include in their sample AGN $\sim80\times$ fainter.  The difference in
these fractions is mainly due to the steep faint-end slope of the
luminosity function at low redshifts \citep{sh06}.  Here, we are
mainly interested in the relative change of the AGN fraction with
galaxy mass.  Clearly, we see that the fraction of galaxies that host
AGN monotonically increases with stellar mass similar to results based
on obscured \citep{ka03b} and radio-loud \citep{be05} AGN in the SDSS.

\subsection{Star formation rates}

In Figure~\ref{sfr1}, we show the SFRs for 1820 zCOSMOS galaxies
(Table~\ref{sample}: Sample B\footnote{The sample is slightly smaller
than that given in Table~\ref{sample}; four AGNs have [OII] falling
outside the observed spectral window that can vary slightly from
slit-to-slit.}) including 43 of them that host moderate-luminosity AGN
($42.0<log~L_{0.5-10.0~{\rm keV}}<43.7$).  The results are presented
before (panels a and b) and after (panels c and d) the correction for
the AGN contribution to the [OII] luminosity based on a single-value
(0.21) of the observed, median [OII]/[OIII] ratio of type 1 AGNs in
the SDSS.  First of all, we see that 88\% of AGNs have detectable
[OII] emission\footnote{The fraction of AGN with detected [OII]
emission is 68\% ($S/N>2$) and 61\% ($S/N>3$) depending on the given
line significance.  We have confirmed our results based on the lower
S/N ratio (1.15) with those based on these smaller, high confidence
samples of AGN.} (panel $a$).  It is apparent by comparing the SFR
distribution for AGN hosts in panels $b$ and $d$ that a significant
shift in the distribution occurs when removing the AGN contribution to
the [OII] emission line.  Even so, we find that the SFRs of AGN host
galaxies are almost exclusively between 1-100 M$_\sun$ yr$^{-1}$
(Fig.~\ref{sfr1}$c$,$d$), a range consistent with analytic models of
AGN hosts with star-forming disks \citep{bal08} and supportive of
constraints on the cosmic IR background \citep{bal07}.

\begin{figure*}
\epsscale{0.9}
\plotone{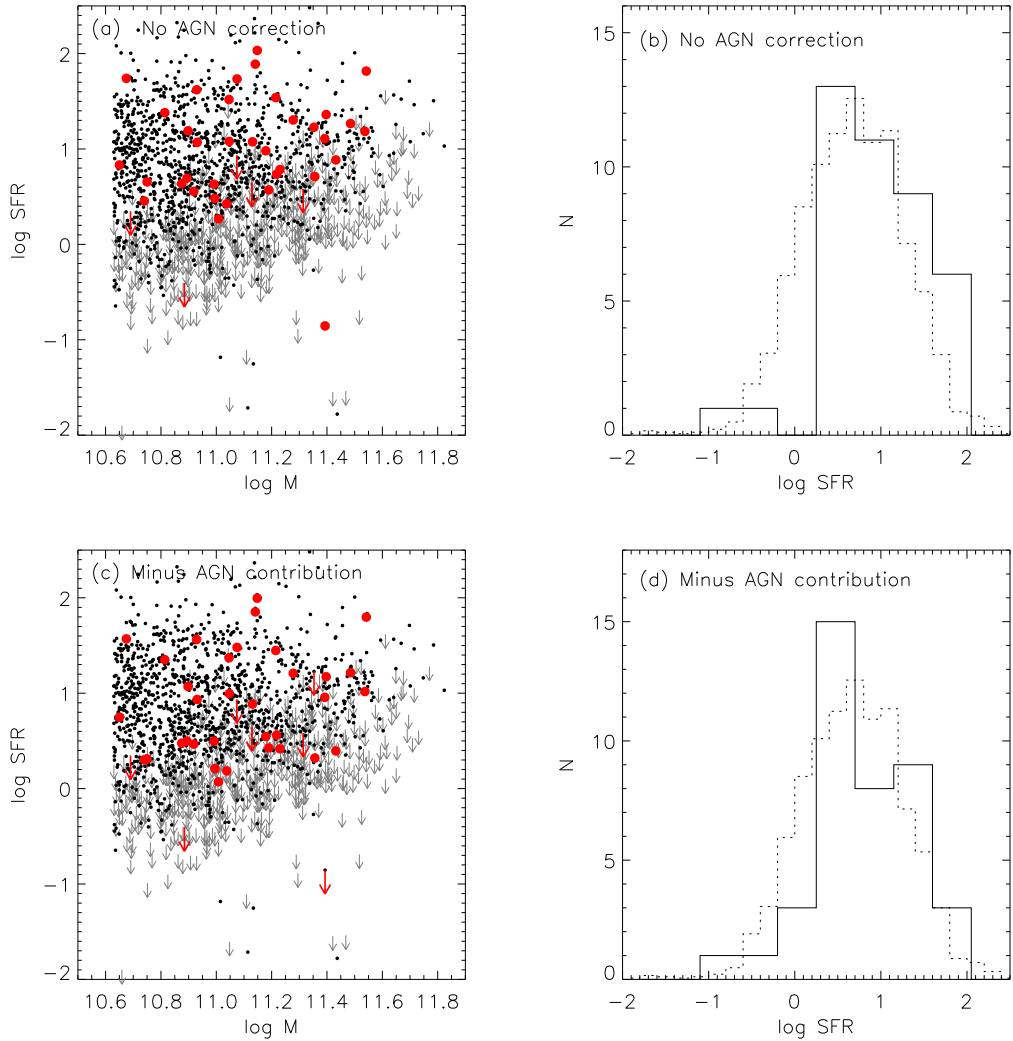}

\caption{Star formation rates ($M_{\sun}$ yr$^{-1}$) versus stellar
mass for galaxies ($0.48<z<1.02$; $log~M_*>10.6$) before (a) and after
(c) the removal of the AGN contribution.  Measurements are shown by
either a small black (emission-line galaxy) or large red (AGN;
$42.0<log~L_{0.5-10~{\rm keV}}<43.7$) circle.  Upper limits are shown
as an arrow with their color descriptive of the type as given above.
($b$, $d$) For ease of comparison, the distributions, including those
with upper limits, are shown by either a solid (AGN hosts) or dashed
(all galaxies) histogram with the full galaxy sample scaled down to
match the AGNs.}

\label{sfr1}
\end{figure*}

To quantitatively test whether the SFR distribution of AGN hosts
differs from that of the underlying galaxy population, we have
determined the SFR distribution, including those with upper limits,
while also considering the dispersion in the AGN contribution to the
observed [OII] emission-line luminosity.  We assume a log normal
distribution of the [OII]/[OIII] ratio with parameters given in
Section~\ref{sfr_measure}.  Based on one thousand iterations, we
determine the mean SFR distribution and perform statistical tests on
each individual iteration compared to the parent galaxy population.
The results of this exercise are shown in Figure~\ref{sfr2} for two
AGN samples of differing luminosities.  First, we consider AGN, as
done above, that fall within the luminosity interval $42<log~L_X<43.7$
for which we have high confidence in our well-matched, mass-selected
parent sample of galaxies.  Based on this limited sample of AGNs, we
cannot significantly discriminate between the two distributions
(Fig.~\ref{sfr2}$a$) given the results of the individual K-S tests
(panel $b$).  Although, we do find that the median SFRs shown in panel
$c$ are for the most part higher than that of the full galaxy sample
(4.7 M$_\sun$ yr$^{-1}$) in each iteration.

We are able to use a less restrictive sample that includes higher
luminosity AGN ($log~L_X>43.7$) thus improving our statistics although
at the expense of maintaining reliable stellar masses across our
sample.  In Figure~\ref{sfr2}$d-f$, we present the equivalent analysis
for AGNs with $log~L_X>42$.  Based on this larger sample, we see that
the SFR distribution of AGN hosts is shifted to higher values than the
overall population (panel $d$).  The K-S tests on the individual
distributions now reject the null hypothesis (i.e., distributions are
equivalent) at the $\sim99\%$ level ($>2.5\sigma$) in essentially all
iterations (panel $e$).  Moreover, the median SFRs are all
systematically higher ($SFR\sim9.5$ M$_\sun$ yr$^{-1}$) than that of
the full galaxy sample.  We further point out that the SFR
distribution of AGN hosts while being elevated in comparison to the
underlying massive galaxy population is essentially equivalent to
those forming stars (i.e., emission-line galaxies).  Therefore, we
conclude that a plentiful gas reservoir is a necessary ingredient for
the fueling of AGN as indicated by the presence of significant star
formation with rates reaching up to $\sim100$ M$_\sun$ yr$^{-1}$.

\begin{figure*}

\includegraphics[angle=90,scale=0.75]{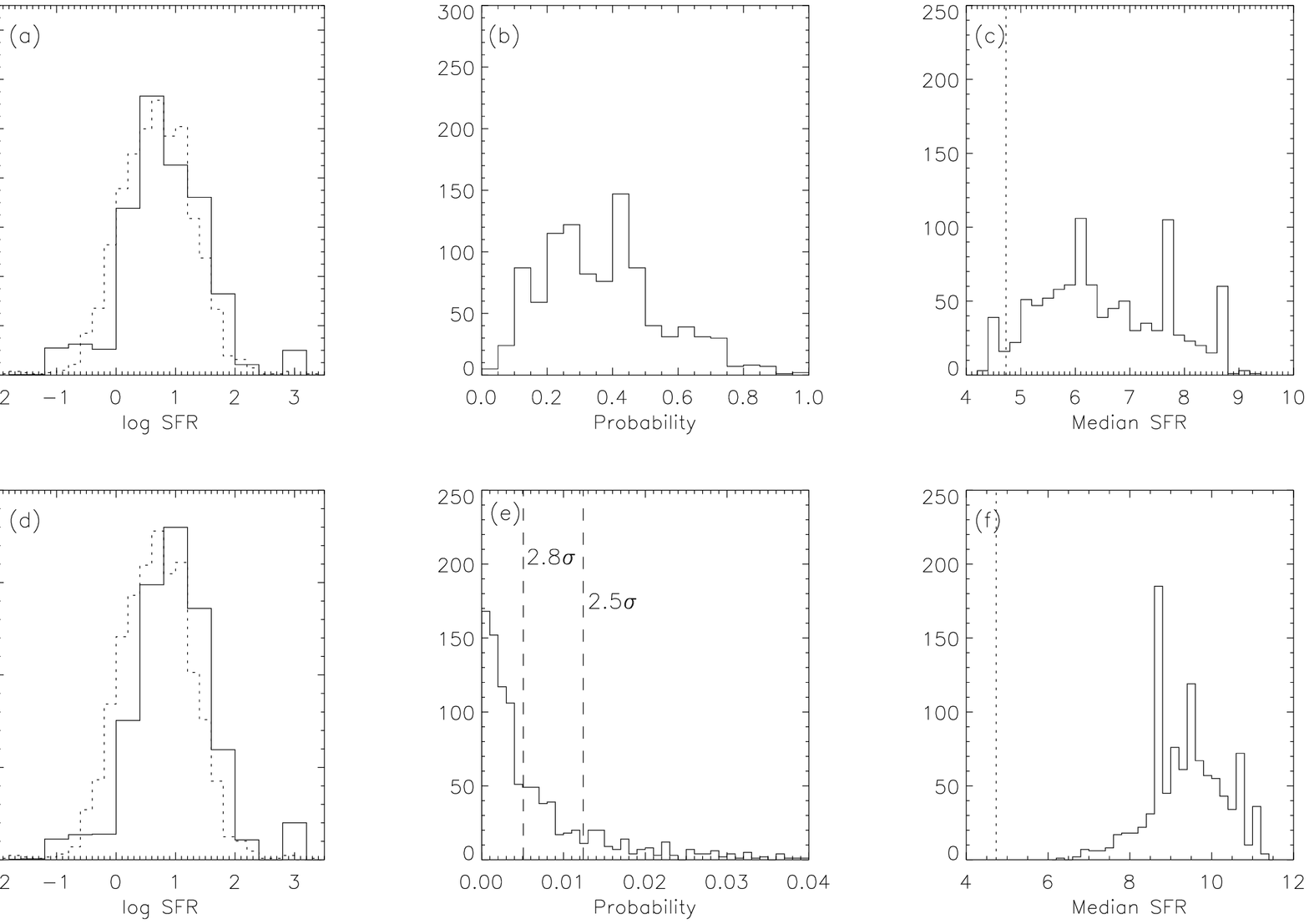}

\caption{Star formation rates of AGN hosts with dispersion in the
applied correction for the AGN contribution.  The two rows are given
for AGN spanning a different luminosity range: ($a-c$)
$42<log~L_X<43.7$, ($d-f$) $log~L_X>42$.  The left panels ($a$, $d$),
equivalent to those in Figure~\ref{sfr1}, show the mean distribution
of SFR, including upper limits, based on 1000 iterations.  The middle
panels ($b$, $e$) give the probability distribution based on a KS
tests that the two samples are equivalent.  The right panels ($c$,
$d$) display the distribution of the median SFR for each iteration
with the median SFR (4.73 $M_{\sun}$ yr$^{-1}$) of all galaxies
depicted by the vertical dashed line.}

\label{sfr2}
\end{figure*}

We investigate how star formation in galaxies hosting AGN is evolving
compared to that of the parent galaxy population.  In
Figure~\ref{sfr_evol}, we show SFR as a function of redshift for
galaxies with $log~M>10.6$ and those hosting AGN (Table~\ref{sample};
Sample B).  A systematic shift of the distribution of zCOSMOS galaxies
towards higher SFRs with increasing redshift is clearly evident with
those hosting AGN exhibiting a similar behavior.  Therefore, we find
that the SFRs of AGN hosts are dictated by that of the underlying
galaxy population thus solidifying similar evidence based on the color
evolution of AGN hosts in the E-CDF-S \citep{si08a}.  As a
consequence, ongoing star formation is most likely responsible for the
significant population of AGN hosts over the redshift range
$0.5\lesssim z \lesssim 1.4$ having blue rest-frame colors
\citep{sa04,bo07,na07} with evidence for such a trend remaining in
place for quasar hosts at higher redshifts $1.8<z<2.8$ \citep{ja04b}.

It is illuminating to compare the SFRs from zCOSMOS to AGN at lower
redshifts.  For consistency with zCOSMOS, we measure SFRs of the type
2 AGN from SDSS \citep[$z<0.3$;][]{ka03b} using the [OII]$\lambda3727$
and [OIII]$\lambda5007$ emission line fluxes from \citet{br04} to
determine an AGN-corrected [OII] luminosity.  We further select only
those galaxies having stellar masses above our zCOSMOS limit
($log~M_*>10.6$) based on the conversion factor (Kroupa to Salpeter
IMF) given in \citet{br04}.  Here, we use the calibration of
\citet{ke04} to derive SFRs\footnote{We have confirmed that the
\citet{mo06} relation gives the same results.} in order to have
consistency with the SFRs measured for the type 1 SDSS sample
\citep{ki06}.  Initially, we demonstrate that the ongoing SFRs shown
in Figure~\ref{sfr_evol} are in agreement with the findings of
\citet{ka03b} based on $D_n(4000)$: weak AGNs ($L_{[OIII]}< 10^7
L_{\sun}$; blue points) have low SFRs ($\sim$0.2 M$_\sun$ yr$^{-1}$)
while strong AGN ($L_{[OIII]}> 10^7 L_{\sun}$; small red points) are
more actively forming stars ($\sim$1 M$_\sun$ yr$^{-1}$).  For ease of
visualization, we show the best-fit relation $log~SFR \propto
log(1+z)$ for only the zCOSMOS galaxies with significant [OII]
emission (upper limits are not considered; black curve) and those
hosting AGN (red curve).  Remarkably, we further find that the SDSS
AGN may be the low redshift analogs of the AGNs in the zCOSMOS survey
given their close proximity to an extrapolation of the evolution of
zCOSMOS galaxies.  In addition, the low-to-moderate levels of star
formation ($\approx 0.5-3$ M$_\sun$ yr$^{-1}$) in type 1 SDSS AGN
\citep[shown roughly by the green triangle in
Fig.~\ref{sfr_evol};][]{ki06} are in agreement with the type 2 AGN
from SDSS and the aforementioned passive evolutionary scenario.  In
light of these results from the SDSS that effectively extend our
redshift baseline, we reiterate our conclusion that the SFRs of AGN
host galaxies are reflective of the overall star formation history of
galaxies and provide no indication of the suppression or truncation
due to a mechanism related to the AGN itself.

\begin{figure*}
\epsscale{0.8}
\plotone{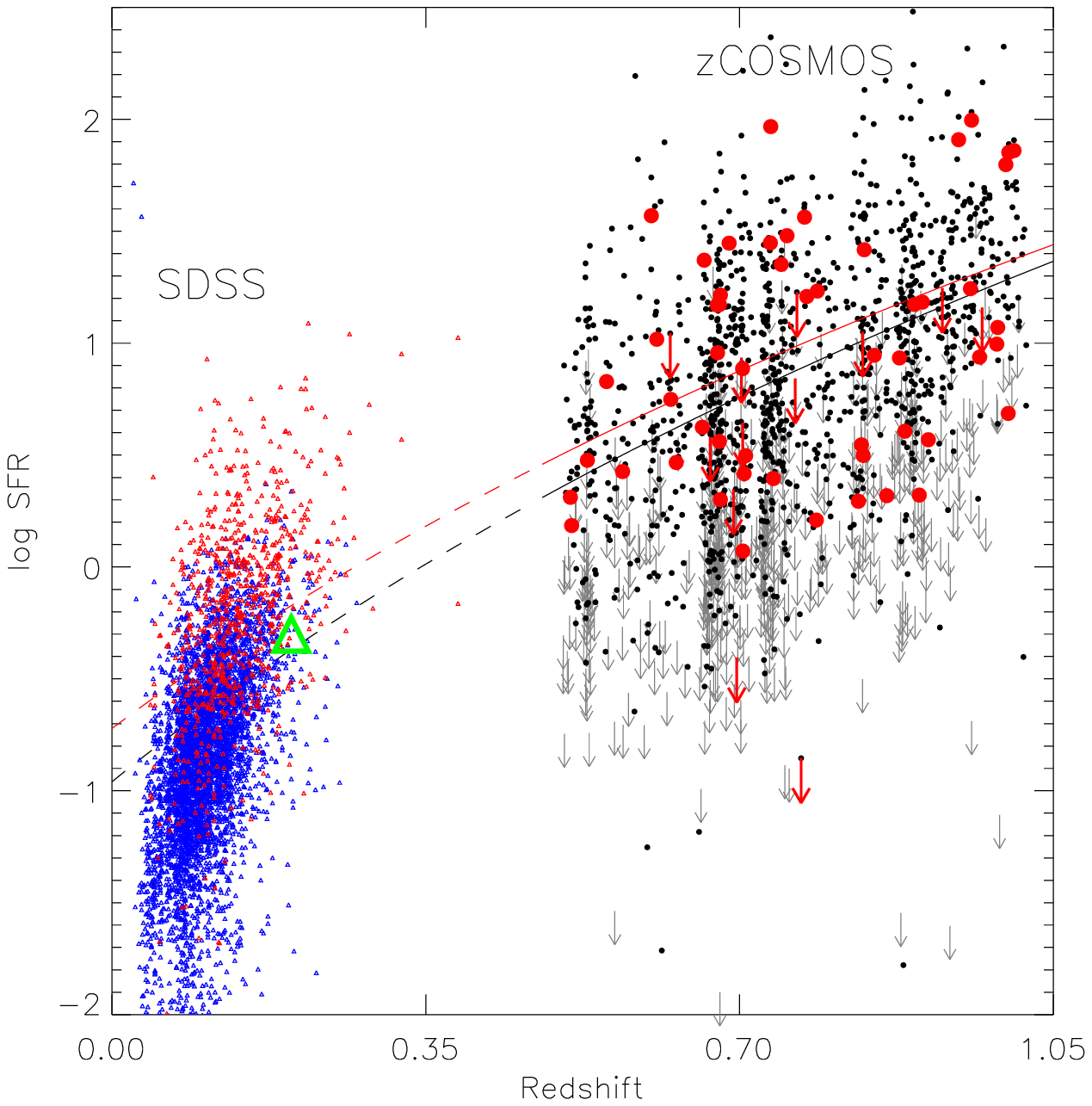}

\caption{Cosmic evolution of star formation.  At $z>0.48$, we show the
SFR-$z$ distribution for all zCOSMOS galaxies with $log~M>10.6$ (small
black circles and crosses) and those hosting AGN with $log ~L_X>42$
(73; large red circles and arrows).  The best-fit linear relation for
the emission-line objects is shown for both zCOSMOS populations
(black: galaxies; red: AGN hosts) with an extrapolation to lower
redshifts (dashed lines).  For comparison, we plot SFRs of AGN hosts
from the SDSS with an equivalent selection on stellar mass; obscured
AGN (type 2) from the sample of \citet{ka03b} are shown with strong
AGN ($log~L_{OIII}>40.5$) in red and those of lower luminosity in
blue.  A large green triangle marks the mean value of the SFR for SDSS
type 1 AGN \citep{ki06}.}

\label{sfr_evol}
\end{figure*}

\subsection{Stellar ages}

To complement our study of the ongoing SFRs of AGN hosts, we use the
spectral index $D_n(4000)$ \citep{ba99}, determined by our
"platefit\_vimos" routine for each galaxy in the zCOSMOS sample to
infer the age of the overall stellar population on longer timescales
($>0.1$ Gyr).  This index is the ratio of the average flux density
$F_{\lambda}$ in the continuum bands 3850-3950 {\rm \AA} and
4000-4100 {\rm \AA}.  This is essentially a measure of the strength
of the 4000-{\rm \AA} break with galaxies having experienced a
recent episode of star formation exhibiting a smaller index due to the
presence of young stars.  In Figure~\ref{age}, we show the values of
$D_n(4000)$ for galaxies with respect to their specific SFR. We
implement two mass limits ($log~M>10.6$: panels $a$, $b$;
$log~M>11.1$: panels $c$, $d$) in order to check for consistency when
the underlying $D_n(4000)$ distribution of galaxies is significantly
different: the higher mass cut results in a distribution dominated by
more evolved galaxies as evident by its peak at $D_n(4000)\sim1.8$.
Initially, it is apparent (Fig.~\ref{age}$a$) that there is a good
correspondence between $D_n(4000)$ and specific SFR for all galaxies
including those with AGNs.  We then use the relation given in
\citet{ka03a} that assumes an instantaneous burst model of
solar-metallicity to infer stellar age from the value of $D_N(4000)$
as given by the scale bar in Figure~\ref{age}a; this relation provides
us with a rough assessment of the actual ages since many of the
zCOSMOS galaxies including those with AGN may have had a more sedative
existence in their recent past.  As a result, we find that most
galaxies with AGN contain a young stellar component since 70\% have
$D_n(4000)<1.6$ ($age \lesssim 2~ Gyr$; Fig~\ref{age}a, c).
Furthermore, we measure the fraction of galaxies hosting an AGN as a
function of $D_n(4000)$ and demonstrate (Fig.~\ref{age}$b$, $d$) that
a galaxy is more likely to have an accreting SMBH if there is a
sufficient supply of gas, fully consistent with the mass measurements
of HI \citep{ho08} and CO \citep{sco03} in AGN hosts, given the higher
rate of occurance in galaxies with younger ages.  These results
effectively extend such studies based on $D_n(4000)$ at low redshift
\citep{ka03b} up to $z\sim1$.

\begin{figure*}
\epsscale{1.0}

\plotone{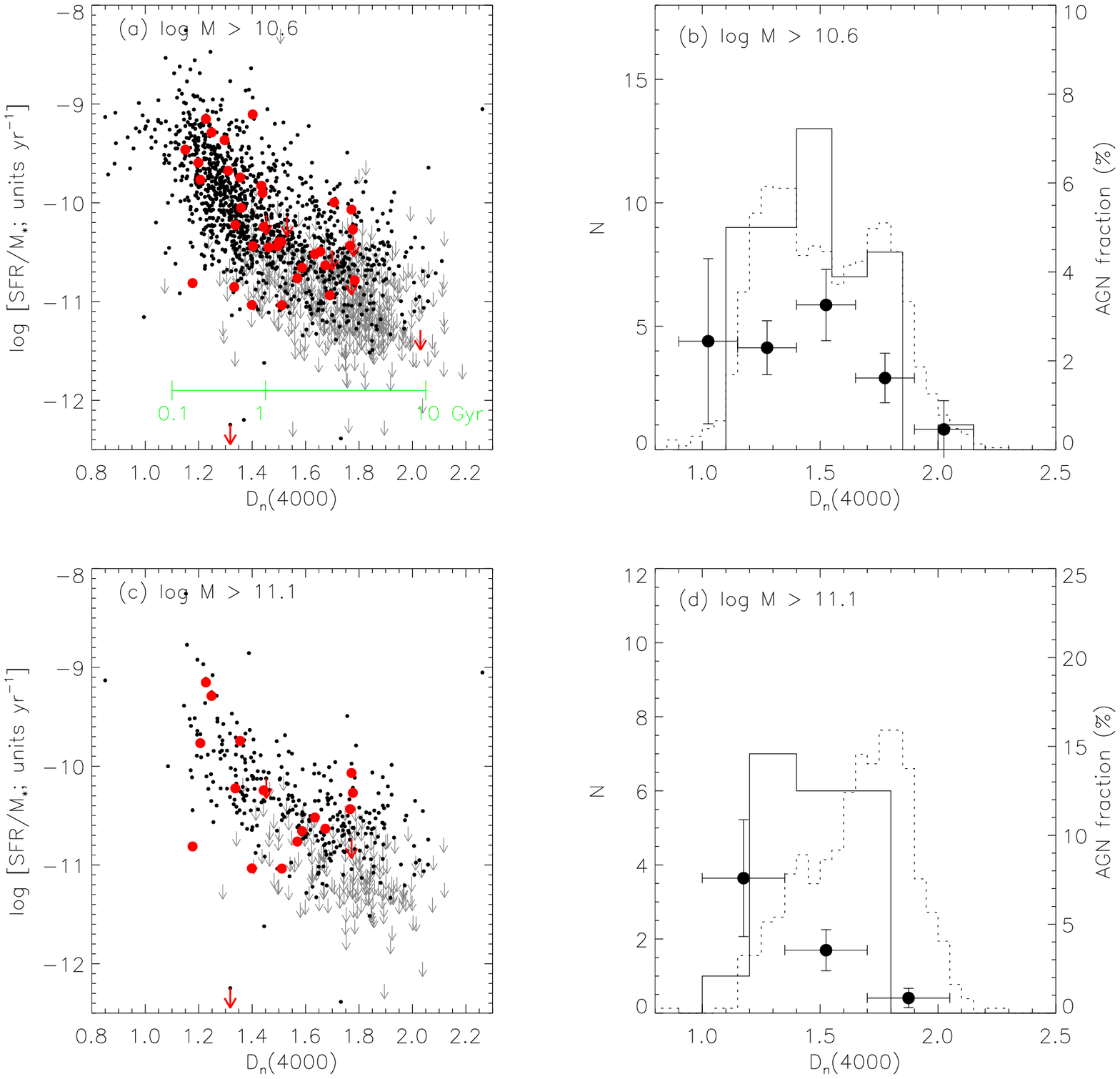} 

\caption{($a$) Specific SFR versus the stellar age indicator
D$_n$(4000).  Symbols are equivalent to those in Figure~\ref{sfr1}a.
An approximate age scale is given based on Figure 2 of
\citet{ka03a}. ($b$) Number distribution and the fraction of galaxies
hosting an AGN as a function of $D_n(4000)$.  (c, d) The equivalent
plots are shown for a higher mass cut as given.  AGN activity is
primarily associated with young stellar populations over a broad range
in host galaxy mass.}

\label{age} 
\end{figure*}

There is a noteable discrepancy between our claim for a higher level
of AGN activity for star-forming galaxies and the location of AGN
hosts on the color-magnitude diagram.  Both X-ray selected
\citep{na07,si08a} and optically-selected AGN \citep{ma07} exhibit a
low AGN fraction along the red sequence, enhanced activity in the
intermediate region (i.e., "green valley") between the blue and red
galaxy populations, and a dropoff towards very blue galaxies.  On the
contrary, the mass-selected AGN sample of \citet{ka03b} clearly have
stellar properties very similar to late-type galaxies (see their
Fig. 14) and, thus a low fraction of AGN within the 'blue cloud'
reported by the aforementioned studies is surprising.  Interestingly,
\citet{le08} find that the fraction of the stacked X-ray signal of
late-type galaxies attributed to AGN emission continuously rises with
SFR.

To investigate this further, we measure the fraction of galaxies
hosting AGN as a function of their rest-frame optical color ($U-V$)
for both a luminosity and mass-selected sample
(Figure~\ref{color_mag_mass}$a$, $b$).  Since we are not constrained
by the use of [OII] for this exercise, we have extended the redshift
baseline $0.1<z<1.02$ to improve the statistics.  In the first panel,
we find that the fraction of galaxies with X-ray selected AGN is
strongly peaked in the "green valley" for the luminosity-selected
sample (panel a) thus agreeing with the results of \citet{si08a}.  In
contrast, the mass-selected sample, behaves similarly for colors
$U-V>1.4$, but does not have a decline towards bluer colors $U-V<1.4$.
This is easily understood since blue galaxies are known to have lower
mass-to-light ratios and the AGN fraction rises substantially with
host galaxy mass as demonstrated in Section~\ref{mass}.  We conclude
that blue, star forming galaxies do have enhanced levels of AGN
activity and there is no evidence for a significant delay in the
emergence of nuclear activity with respect to the onset of star
formation.  This result is consistent with the recent findings of
\citet{si08a}, based the morphology of the host galaxy, that blue
($U-V<0.7$), bulge-dominated ($n_{sersic}>2.5$) galaxies have the
highest incidence (21.3\%) of AGN activity, and the blue rest-frame
colors of low redshift ($z<0.2$) quasars irrespective of their
morphology \citep{ja04a}.  However, there remains the possibility that
the lower fraction of red galaxies hosting AGN is due to some form of
self-regulating feedback that effectively reduces the AGN luminosity
below our flux limit.  Nonetheless, these results lend no support for
a simple model prescription that attributes the truncation of star
formation or the color evolution of galaxies to AGN feedback and bring
into question whether studies based on the optical emission line
diagostics of AGN activity miss a significant number of those residing
in star-forming galaxies \citep[see][for a discussion on this
topic]{sch07}.

\begin{figure*}
\epsscale{1.0}

\plotone{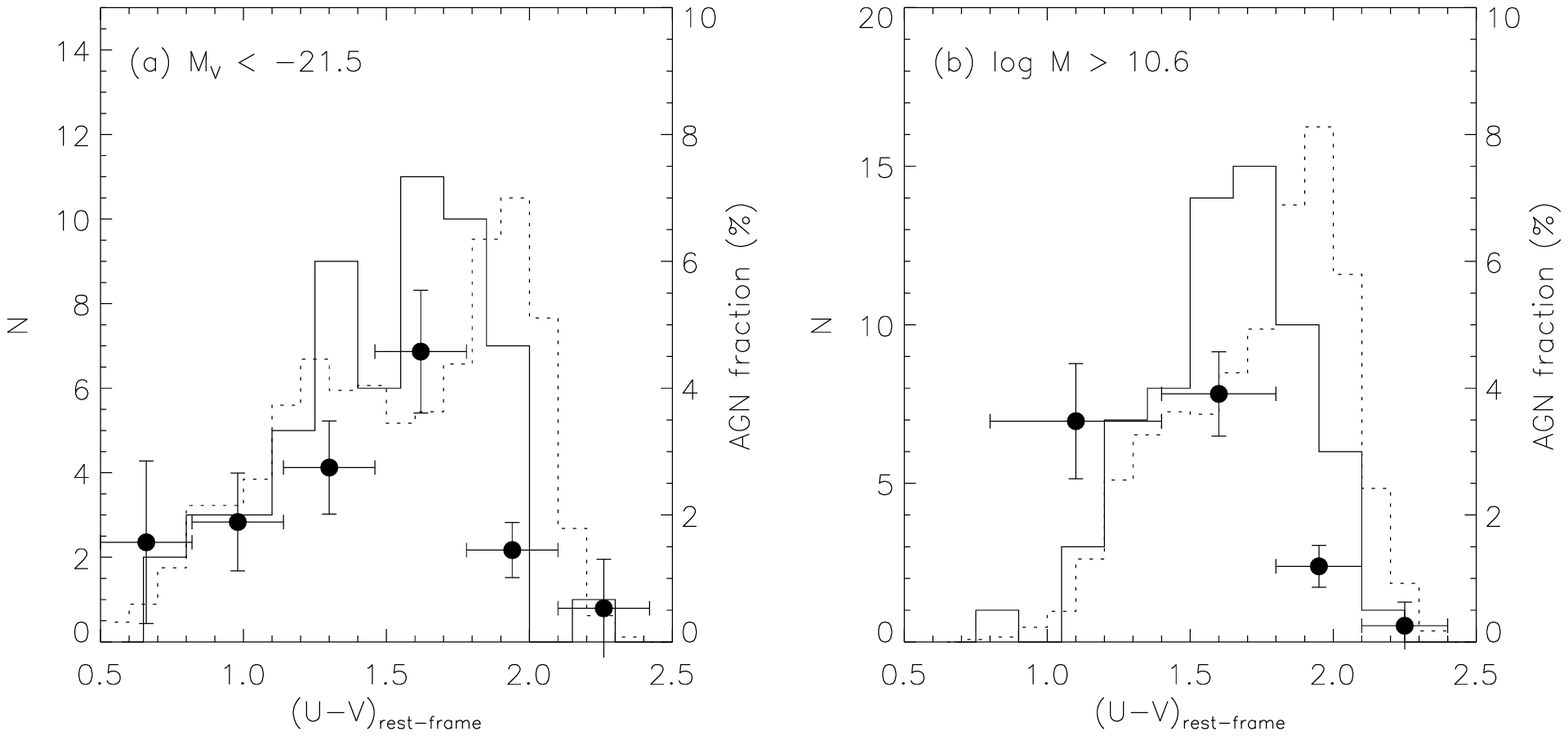}

\caption{Rest-frame color distribution $U-V$ of galaxies
($0.1<z<1.02$) and the fraction hosting AGN.  We show the distribution
for 58 galaxies (solid histogram) selected by optical luminosity
($M_V<-21.5$; panel a) and 65 galaxies above a fixed mass limit
($log~M>10.6$; panel b) that host AGN.  For comparison, the
distribution for the parent galaxy population (dashed histogram) is
normalized to match the AGNs in each panel.  The decline in the AGN
fraction towards the bluest colors (panel a) appears to be due to the
inclusion of galaxies with low mass-to-light ratios that are not
present in the mass-selected sample (panel b).}

\label{color_mag_mass}
\end{figure*}

\section{AGN-star formation connection and co-evolution}

Having established an association between AGN activity and star
formation, we are motivated to determine how closely these phenomena
are related on a case-by-case basis that can potentially signify an
underlying causal connection.  The existence of such a relationship
may be realized given the recent findings that AGN accretion power as
probed by [OIII] luminosity is higher for galaxies with younger
stellar populations \citep{ka07}.  Ideally, we would like to
investigate such a relation using a quantity more closely associated
with the accretion process, namely X-ray emission, known to originate
closer to the black hole and an indicator of the instantaneous star
formation rate.

To do so, we have plot in Figure~\ref{agn_gal_relations}a the SFRs of
our AGN sample, as measured by [OII] strength, versus hard X-ray
luminosity (2-10 keV) and inferred mass accretion rate while
implementing a bolometric correction as given in \citet{ma04} and an
accretion efficiency of 0.1.  We find that a weak correlation exists
based upon a Pearson correlation coefficient of 0.17 and a linear fit
that has a shallow slope with significant dispersion in its value
($0.28\pm0.22$).  We conclude that underlying complexities such as the
efficiency of transferring gas to nuclear region over kiloparsec
scales, and varying duty cycles for star formation and accretion may
contribute to the large dispersion in these relations.

\begin{figure}

\plotone{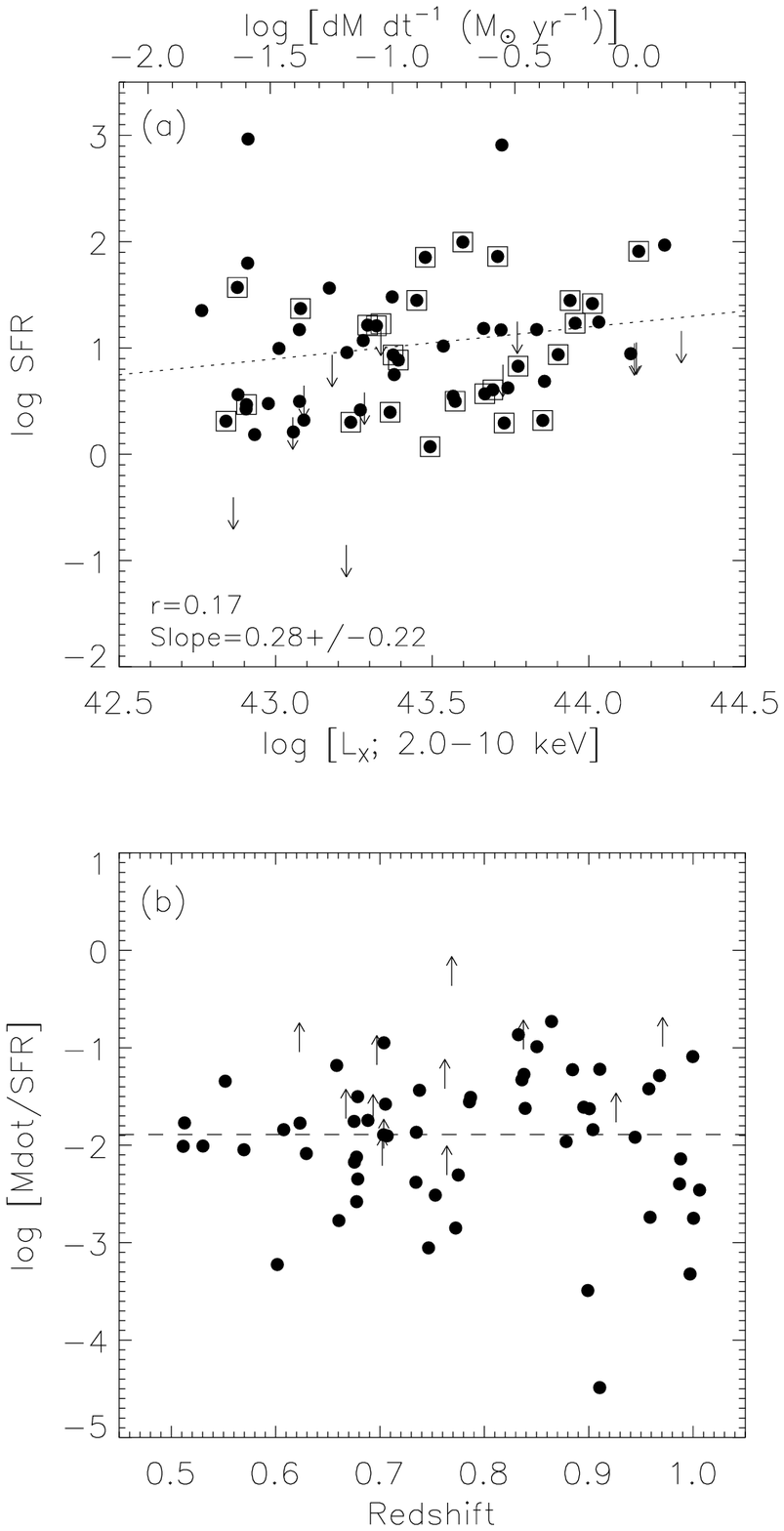}

\caption{AGN-galaxy relations: ($a$) SFR versus hard X-ray luminosity
and mass accretion rate including the best fit relation.  Absorbed AGN
with detected star formation are further highlighted by an open box.
($b$) Ratio of mass accretion to SFR versus redshift.  The horizontal
dashed line marks the median ratio.  Measurements are shown by
a solid circle in both panels while limits either upper ($a$) or lower
($b$) are given by an arrow.}

\label{agn_gal_relations} 

\end{figure}

There have been some recent claims that an obscured phase, coupled
with enhanced star formation, may represent an early stage in the
subsequent evolution of AGN
\citep[e.g.,][]{pa04,al05,ki06,hop08a,po08}.  To test this scenario,
we have marked those AGN in Figure~\ref{agn_gal_relations}a that have
excessive X-ray absorption ($N_{\rm H} \gtrsim 10^{22}$ cm$^{-2}$)
based on their hardness ratio [$HR=(H-S)/H+S)> -0.2$] determined by
the X-ray counts in the soft (S: 0.5-2.0 keV) and hard (H: 2-10 keV)
bands assuming a powerlaw spectrum with a photon index of 1.9 at an
effective redshift of $z\sim0.7$.  We find that the absorbed sources
span the same range of SFR as the unabsorbed sources.  A KS test give
a probability of 65\% that both absorbed and unabsorbed AGN could be
drawn from the same parent population.  These results are not
dependent on the chosen division in hardness ratio.  Therefore, we
conclude that an amendment, such as the aforementioned evolutionary
scenario, to the unification model \citep{an93} for these
moderate-luminosity AGNs is not supported by our findings.  Although,
there remains substantive evidence that enhanced star formation may be
associated with nuclear obscuration for the more luminous QSOs
\citep[e.g.,][]{pa04,la07,za08} possibly due to different physical
mechanisms for triggering mass accretion on SMBHs.

It is of much interest to determine the relative growth rate of a
galaxy and its central SMBH as a function of redshift in light of the
well-established local relation
$M_{BH}/M_{bulge}\approx1.5\times10^{-3}$ \citep[e.g.,][]{mc02,ha04}.
We plot in Figure~\ref{agn_gal_relations}b the ratio of mass accretion
rate onto a SMBH to the SFR as a function of redshift.  The median
value ($1.9\times10^{-2}$) is roughly an order-of-magnitude higher
than the local ratio $M_{BH}/M_{bulge}$.  This difference is likely
due to the varying timescales between SMBH accretion and star
formation $\Delta t_{BH}/\Delta t_{gal}\approx M_{BH}/M_{bulge} \times
SFR/dM_{accr} dt^{-1}\approx 0.1$.  We note that our SFR measurements
most likely include a significant disk component thus a more rigorous
assessment is required and beyond the scope of this work.
Nonetheless, by assuming that star formation occurs over rougly a
dynamical timescale $\Delta t_{gal}\approx10^9$ yr, the duty cycle for
AGN activity ($\approx10^8$ yr) is consistent with the
luminosity-dependent model predictions of \citet{hop08b} for SMBHs
with $M_{BH}\sim10^8$ M$_{\sun}$ and accreting above an Eddington
ratio of 0.01, both within a physical regime spanned by our sample
assuming these AGNs have already settled on a SMBH - bulge relation at
their respective redshifts.  Alternatively, if we assume a timescale
for galaxy growth to be the inverse of their sSFRs as shown in
Figure~\ref{age}, the AGN lifetimes can reach up to $\sim10^{10}$ yr
that fully illustrates that these AGNs and their host galaxies are
close to being fully matured although have SMBH growth times at least
an order-of-magnitude less than those in the SDSS \citep{he04}.
Furthermore, we find that there is no dependence of the ratio
($dM_{accr} dt^{-1}/SFR$) on redshift thus supporting a
co-evolutionary scenario where both the average SFR and mass accretion
rates onto SMBHs are rapidly declining with equivalent rates from
$z\sim1$ to the present possibly due to diminishing fuel supplies.

These comparisons suggest that beyond the local universe the $M_{BH} -
M_{bulge}$ relation should display higher intrinsic dispersion given
the large spread over two orders of magnitude ($10^{-3}-10^{-1}$) in
the relation between accretion rate and SFR
(Fig.~\ref{agn_gal_relations}$b$).  \citet{ro06} point out that AGN
samples at high redshift may have larger intrinsic scatter in the
velocity dispersion of their hosts due to their immature dynamical
state. The observational situation is so far unclear: limited AGN
samples at higher redshifts \citep{wo06,wo08} show preferentially
higher black hole masses relative to that expected by the local $M -
\sigma$ relation, while other studies of low redshift AGNs, undergoing
substantial rates of accretion ($L_{Bol}/L_{Edd}>0.1$), have black
hole masses lower than that of inactive galaxies of equivalent host
mass \citep{ho08} or luminosity \citep{ki08}.  Notwithstanding the
large observational biases \citep{lau07}, it is possible that the
coeval growth of galaxies and their SMBHs may be intermittent with
either stars or SMBHs growing somewhat faster or slower than each
other, as indicated by our findings exemplified in
Figure~\ref{agn_gal_relations}$b$, while still resulting in a tight
black hole-bulge mass relation for inactive galaxies at $z=0$ as
suggested by \citet{wo06}.

\section{Summary and conclusions}

We have utilized the zCOSMOS 10k catalog of galaxies with reliable
spectroscopic redshifts to investigate the properties of those that
host AGN.  X-ray observations with $XMM$-Newton enable us to identify
152 AGN, from a parent sample of 7543 zCOSMOS galaxies, that include
those with significant obscuration and of low optical luminosity.  The
derived properties of the full zCOSMOS sample such as stellar mass,
rest-frame colors, and spectral properties (i.e., emission line
strength, $D_n4000$) enable us to determine the prevalence of AGN
activity as a function of galaxies with the aforementioned
characteristics.

Specifically, we measure the SFR of galaxies using the
[OII]$\lambda$3727 line luminosity.  We account for the contribution
from the underlying AGN component most likely arising from the narrow
line region by using the [OIII]$\lambda$5007 luminosity and the
typical [OII]/[OIII] ratio found from previous studies of AGN and
quasars \citep{ho05,ki06}.  The [OIII] line luminosity is measured
directly from our spectra if present.  For the subsample with [OIII]
outside our observed spectral bandpass, we infer the [OIII] strength
from the hard (2--10 keV) X-ray luminosity and the well known
correlation between these two quantities.

Overall, we find that the two main requirements for a galaxy to host
an actively, accreting SMBH are (1) its stellar mass and (2) a
significant amount of gas content as inferred by the observed levels
of star formation and the age of their stellar populations.  These
findings essentially extend those of the SDSS \citep{ka03b} out to
$z\sim1$.

We particularly draw the following conclusions:

\begin{itemize}

\item We confirm with many previous studies that the fraction of
galaxies hosting AGN rises with increasing mass with most host
galaxies having $M_*>4\times10^{10}$ $M_{\sun}$.

\item The host galaxies of AGN have ongoing star formation with a
broad range of rates ($\approx1-100$ $M_{\sun}$ yr$^{-1}$) higher than
that of the overall massive ($log~M>10.6$) galaxy population and
essentially equivalent to those forming stars (i.e., emission-line
galaxies).  The association of AGN activity and young stellar
populations is further substantiated by an observed increase in the
fraction of galaxies harboring AGN at low values of the spectral index
$D_n(4000)$ and blue rest-frame color $U-V$.

\item The enhancement of AGN in young, star-forming galaxies with
$1.0<D_n(4000)<1.4$ and $0.8<U-V<1.5$ lessens the evidence for AGN
prefering to reside in galaxies of intermediate colors and their role
in galaxy evolution.  We demonstrate that previous studies based on
luminosity-selected samples are misleading due to the inclusion of
galaxies with low mass-to-light ratios that are less likely to harbor
AGN given their lower masses.

\item The SFRs of AGN hosts evolves with redshift in an equivalent
manner to the overall star-forming galaxy population.  This
essentially brings the evidence for a co-evolution scenario between
accretion onto SMBHs and the star-formation history of galaxies to a
closer physical scale (i.e., within the same galaxies).

\item A direct relationship between the consumption of gas into stars
and that accreted onto SMBHs is weak, suggesting that additional
physical complexities or varying timescales may be inherent.  On
average, a co-evolution scenario for the overall population is clearly
evident given the constancy of the ratio ($\sim10^{-2}$) between mass
accretion rate onto SMBHs and SFR with redshift possibly indicative of
depleting gas reservoirs from $z\sim1$ to the present.  The
order-of-magnitude increase in this ratio compared to the locally
measured value of $M_{BH}/M_{bulge}$, is consistent with an AGN
lifetime substantially shorter than that of star formation.
Furthermore, the significant dispersion in this ratio may be
indicative of larger scatter in the BH-bulge relations at higher
redshift.

\end{itemize}

Our results put important constraints on physical models of AGNs and
their evolution.  The considerable rates of star formation ($\sim10$
M$_\sun$ yr$^{-1}$) in their hosts, coupled with the lack of
structural signs of galaxy interactions \citep{gr05,ga08} and
timescales for accretion, indicate that a "Seyfert mode" of accretion,
driven by secular processes, is more likely than merger-driven models
for this class of moderate-luminosity AGN.  In particular, there is no
indication of the suppression or truncation of star formation at
levels expected from models implementing AGN feedback.  Therefore, we
find no conclusive evidence that AGNs, undergoing this mode of
accretion, are a key factor in the evolution of galaxies.

\acknowledgments

We are most grateful to the referee for providing a critical review
that significantly improved the paper and to Minjin Kim for allowing
us to use his emission line measurements of SDSS AGNs and fielding
subsequent communications.  This work is fully based on observations
undertaken at the European Southern Observatory (ESO) Very Large
Telescope (VLT) under the Large Program 175.A-0839 (P.I., Simon
Lilly).

{\it Facilities:} \facility{XMM}, \facility{VLT:Melipal (VIMOS)}

\begin{deluxetable}{llccll}
\tabletypesize{\small}
\tablecaption{Galaxy and AGN sample\label{sample}}
\tablehead{\colhead{Sample}&\colhead{Redshift}&\colhead{log mass}&\colhead{\# galaxies}&\colhead{\# AGN}&\colhead{Purpose}\\
&range&\colhead{(M$_*$)}&&\colhead{($log~L_{0.5-10~{\rm keV}}$)}}
\startdata
-&0.1-1.02&-----&7543&152 ($>42.0$)\\
-&0.1-1.02&$>10.6$&2540&105 ($>42.0$)\\
A&0.1-0.50&-----&3356&52 ($42.0-43.7$)&Host galaxy mass distribution\\ 
&&&& 32 (42.48-43.7)&Mass-dependent AGN fraction\\
B&0.48-1.02&$>$10.6&1820&73 ($>42.0$)&SFRs\\
&&&&47 ($42.0-43.7$)&Mass-dependent SFRs\\
\enddata
\end{deluxetable}

\end{document}